\documentclass[lettersize,journal]{IEEEtran}
\usepackage{amsmath,amsfonts}
\usepackage{orcidlink}
\usepackage{algorithmic}
\usepackage{algorithm}
\usepackage{array}
\usepackage[caption=false,font=footnotesize,labelfont=rm,textfont=rm]{subfig}
\usepackage{textcomp}
\usepackage{stfloats}
\usepackage{url}
\usepackage{verbatim}
\usepackage{graphicx}
\usepackage{cite}
\usepackage{xcolor}
\usepackage{multirow,booktabs,amssymb,gensymb,tabularx}
\hypersetup{
	colorlinks=true,
	linkcolor=green,
	citecolor=green
}

\begin{document}

\title{Dual-Branch Knowledge Distillation for Noise-Robust Synthetic Speech Detection}
%

\author{Cunhang Fan\hspace{-1.5mm}$^{~\orcidlink{0000-0001-6318-8803}}$, ~\IEEEmembership{Member, IEEE}, Mingming Ding\hspace{-1.5mm}$^{~\orcidlink{0000-0001-5972-2540}}$, Jianhua Tao\hspace{-1.5mm}$^{~\orcidlink{0000-0002-9344-6428}}$, ~\IEEEmembership{Senior Member, IEEE}, Ruibo Fu\hspace{-1.5mm}$^{~\orcidlink{0000-0001-9598-1881}}$, ~\IEEEmembership{Member, IEEE}, Jiangyan Yi\hspace{-1.5mm}$^{~\orcidlink{0000-0003-2422-4618}}$, ~\IEEEmembership{Member, IEEE}, Zhengqi Wen, Zhao Lv\hspace{-1.5mm}$^{~\orcidlink{0000-0001-9727-366X}}$,  ~\IEEEmembership{Member, IEEE}

\thanks{This work is supported by the STI2030-Major Projects (No.2021ZD0201500), the National Natural Science Foundation of China (NSFC) (No.62101553, No.62201002, No.62322120), Distinguished Youth Foundation of Anhui Scientific Committee (No.2208085J05), Special Fund for Key Program of Science and Technology of Anhui Province (No.202203a07020008), Open Fund of Key Laboratory of Flight Techniques and Flight Safety, CACC (No.FZ2022KF15), the Open Research Projects of Zhejiang Lab (No.2021KH0AB06), Cloud Ginger XR-1. \emph{(Corresponding authors: Jianhua Tao; Ruibo Fu; Zhao Lv.)}
	
Cunhang Fan, Mingming Ding, and Zhao Lv are with the Anhui Province Key Laboratory of Multimodal Cognitive Computation, School of Computer Science and Technology, Anhui University, Hefei 230601, China (e-mail:cunhang.fan@ahu.edu.cn; e21301171@stu.ahu.edu.cn; kjlz@ahu.edu.cn). 

Ruibo Fu, Jiangyan Yi, and Zhengqi Wen are with the NLPR Lab, CASIA, Beijing 100190, China (e-mail: ruibo.fu@nlpr.ia.ac.cn; jiangyan.yi@nlpr.ia.ac.cn; zqwen@nlpr.ia.ac.cn).

Jianhua Tao is with the Department of Automation, Tsinghua University Beijing National Research Center for Information Science and Technology, Tsinghua University (e-mail: jhtao@tsinghua.edu.cn).

}
}

\markboth{Journal of \LaTeX\ Class Files,~Vol.~14, No.~8, August~2021}%
{Shell \MakeLowercase{\textit{et al.}}: A Sample Article Using IEEEtran.cls for IEEE Journals}

\IEEEpubid{0000--0000/00\$00.00~\copyright~2021 IEEE}

\maketitle

\begin{abstract}
Most research in synthetic speech detection (SSD) focuses on improving performance on standard noise-free datasets. However, in actual situations, noise interference is usually present, causing significant performance degradation in SSD systems. To improve noise robustness, this paper proposes a dual-branch knowledge distillation synthetic speech detection (DKDSSD) method. Specifically, a parallel data flow of the clean teacher branch and the noisy student branch is designed, and interactive fusion module and response-based teacher-student paradigms are proposed to guide the training of noisy data from both the data distribution and decision-making perspectives. In the noisy student branch, speech enhancement is introduced initially for denoising, aiming to reduce the interference of strong noise. The proposed interactive fusion combines denoised features and noisy features to mitigate the impact of speech distortion and ensure consistency with the data distribution of the clean branch. The teacher-student paradigm maps the student's decision space to the teacher's decision space, enabling noisy speech to behave similarly to clean speech. Additionally, a joint training method is employed to optimize both branches for achieving global optimality. Experimental results based on multiple datasets demonstrate that the proposed method performs effectively in noisy environments and maintains its performance in cross-dataset experiments. Source code is available at \emph{https://github.com/fchest/DKDSSD}.

\end{abstract}


\section{Introduction}
\IEEEPARstart{S}{peech} synthesis technology has developed rapidly in recent years. High-quality text-to-speech synthesis \cite{tan2021survey} applies advanced deep learning frameworks and has been able to generate synthesis speech that is nearly indistinguishable from real speech. This synthesis speech not only poses a serious security threat to the biometric verification system \cite{wu2015spoofing}, but it may also interfere with the dissemination of internet information. Therefore, to ensure network security, it is crucial to improve the detection capabilities of synthesis speech.

Many studies focus on improving model performance under clean conditions. The ASVspoof series of challenges \cite{wu2015asvspoof, kinnunen2017asvspoof, todisco19_interspeech} has contributed high-quality noise-free datasets. Some works explore discriminative subbands \cite{yang2019significance, 2021The,wang2022low,xue2023learning, ling2021attention} to improve the representation ability of features. Additionally, there are more works \cite{tak2021end, tak2021end2, hua2021towards} that tend to utilize original waveforms for end-to-end detection, thereby avoiding the information loss associated with hand-crafted features. Recent classification networks \cite{li2021replay, jung2022aasist} are predominantly based on convolutional neural network and exhibit strong modeling capabilities for features. Although the above models demonstrate excellent performance on standard datasets, they cannot often generalize to noisy environments.

\begin{figure*}[!t]
	\centering
	\includegraphics[width=\linewidth]{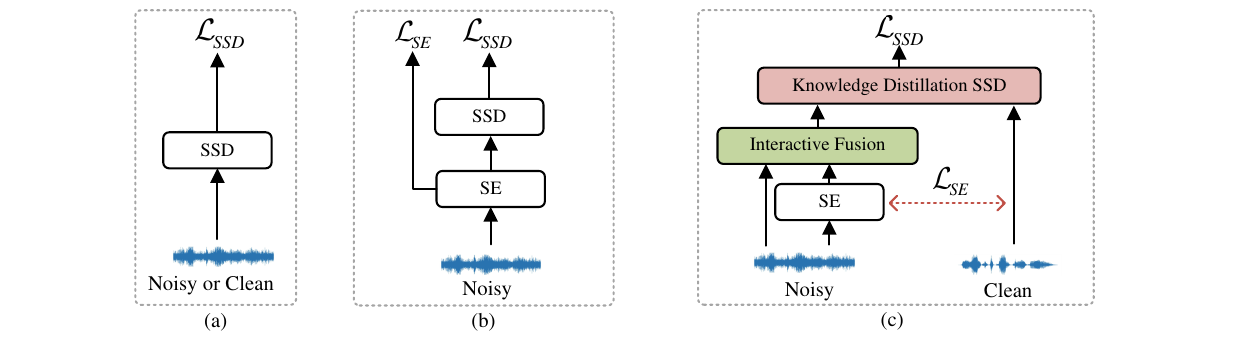}
	\caption{The structure of (a) traditional synthetic speech detection, using clean speech training or using noisy speech for multi-condition training (MCT); (b) cascade or joint training with a speech enhancement front end for noisy scenarios; (c) The overall joint training structure of the proposed dual-branch knowledge distillation synthetic speech detection system. The teacher model requires clean speech as input and is trained in parallel with the student model.}
	\label{fig:structure}
\end{figure*}

\IEEEpubidadjcol
In earlier studies, it has been shown that additive noise \cite{tian2016investigation} can significantly degrade the performance of detectors trained on pure speech, so it is necessary to improve noise robustness. Recently, there have also been related competitions. The ADD 2022 \cite{yi2022add} competition proposes a low-quality fake audio detection track, and the ASVspoof 2021 \cite{yamagishi2021asvspoof} logical access (LA) track also introduces channel noise interference. When dealing with noisy data, data augmentation \cite{cohen2022study, chen2020generalization, zhang2021empirical, yan2022audio} is a widely employed method, which can effectively improve the features representation ability of the model. Multi-condition training\cite{qian2017deep} is based on the principle of data augmentation, utilizing noisy data for training to improve the model's perception of noise (as shown in Fig.~\ref{fig:structure}(a)). In addition, the large-scale speech pre-training model \cite{babu2021xls} has been fine-tuned for the SSD task in \cite{wang2021investigating}. Adversarial training \cite{zhang2021empirical} seeks noise-invariant features by adding a gradient inversion layer into the classifier. However, data augmentation and pre-trained models exhibit limitations in generalizing to unknown noise types, and the use of adversarial strategies may increase training instability.

Another approach is to introduce speech enhancement to learn noise masks \cite{gomez2019gated, dicsken2023differential}, thereby enhancing the acoustic conditions. As illustrated in Fig.~\ref{fig:structure}(b), speech enhancement serves as the front-end, and is cascaded with the SSD model\cite{ma2021multitask}, or the two modules are jointly trained \cite{subramanian2019speech, wang23v_interspeech}. However, the cascade method faces challenges in achieving global optimization, and speech enhancement in the joint training method may lead to unavoidable speech distortion problems.

To improve the performance in noisy scenes, this paper proposes a dual-branch knowledge distillation method for noise-robust synthetic speech detection (DKDSSD). We propose interactive fusion and response-based teacher-student paradigms, design parallel flows for noisy and clean data, and utilize a clean teacher model to guide the training of noisy data. Speech enhancement is applied at the front-end for denoising, but it inevitably introduces speech distortion due to estimation errors. Therefore, it's essential for enhanced features to retain only the information that is useful downstream. The proposed interactive fusion module can adaptively combine the beneficial information of the original noisy features and the denoised features to generate noise-robust features. Knowledge distillation promotes students to learn the classification ability of clean teachers. It involves mapping the decision space of the student model to that of the teacher. Additionally, joint training is used to optimize the entire structure so that the teacher-student network reaches the global optimum. The main contributions of this work are summarized as follows:

\begin{itemize}
	\item[$\bullet$] Knowledge distillation from clean scenes to noisy scenes is proposed, utilizing response-based knowledge distillation loss to constrain the noisy branch to learn the final prediction of the clean teacher.
	\item[$\bullet$] Interactive fusion is proposed to enable the channel interaction of denoised and noisy features, and fuse them at the spatial level, adaptively reducing noise interference and balancing distortion issues.
	\item[$\bullet$] Extensive experiments are conducted on multiple simulated noisy datasets and official datasets, and the results show that our proposed method outperforms the cascade system or joint training method, while maintaining performance on clean scenes. In cross-dataset experiments, the proposed method exhibits the best generalization.
\end{itemize}
	
The rest of this paper is structured as follows. Section \uppercase\expandafter{\romannumeral2} introduces the related work, Section \uppercase\expandafter{\romannumeral3} describes the proposed DKDSSD method, Section \uppercase\expandafter{\romannumeral4} gives the details of the experimental setup, and Section \uppercase\expandafter{\romannumeral5} discusses the experimental results and presents visual analyses. Finally concludes in Section \uppercase\expandafter{\romannumeral6}.

\section{Related Works}
\subsection{Noise Robustness}

Under additional noise and reverberation conditions\cite{tian2016investigation}, SSD trained on clean speech has the problem of insufficient generalization. In \cite{hanilci2016spoofing}, different front-end features and traditional speech enhancement techniques are applied, but they did not significantly improve accuracy. Speech enhancement \cite{zhang2022time, zheng2023sixty} aims to restore a clean spectrum from noisy speech and can be used as a front-end \cite{cai2020within, shon2019voiceid, kinoshita2020improving} to preprocess speech and reduce the interference of strong noise. However, in previous work \cite{fan2022specmnet, fan2020gated, fan2023two}, we find that speech enhancement produces unavoidable estimation errors, and unknown artifacts \cite{iwamoto22_interspeech} may be more harmful than noise. To alleviate this phenomenon, the study \cite{sato2022learning} proposes to switch the input between the enhanced signal and the original signal according to the level of signal-to-noise ratio (SNR) to balance noise and distortion. Some works \cite{pandey2021dual, wang2022wav2vec, zhu2022noise, zorilua2022speaker, hu2022interactive} propose to fuse the two states of speech to reduce the recognition error rate. In the SSD task, the original speech contains artifacts from the synthesis method. Therefore, we propose interactive fusion with original speech, which can preserve artifacts while denoising, thereby generating noise-robust embedding.

\subsection{Knowledge Distillation}
Knowledge distillation \cite{gou2021knowledge} is a prevalent method for compressing models. It can effectively train smaller student models from larger teacher models \cite{xue2023learning}, speeding up inference. Even without the need for compression, knowledge distillation can improve the performance of the student model by transferring knowledge \cite{zhu2019low} from the teacher model. Applying knowledge distillation for continuous learning \cite{li2017learning, kim2021fretal} can enhance domain generalization without compromising the retention of source domain knowledge. In addition, a method based on integral knowledge fusion is proposed in \cite{ren2023generalized} to achieve generalized synthetic speech detection. Recent research \cite{liu2021end} applies knowledge distillation to anti-spoofing detection in noisy environments, transferring feature knowledge from clean teacher to noisy student. In experimental setups involving music and white noise, the model achieved significant performance improvements in high SNR scenarios. However, the experimental results indicate that feature distillation fails in low SNR scenarios, highlighting the risks of forcibly aligning feature spaces. We believe that when the difference between the two feature spaces is relatively large, it is more reasonable to choose the response-based distillation. The proposed interactive fusion method can adaptively combine useful information. This ensures that the distribution space of the input data for the student model remains consistent with that of the teacher model.

\begin{figure*}[t]
	\centering
	\setlength{\abovecaptionskip}{-0.1cm}
	\includegraphics[width=\linewidth]{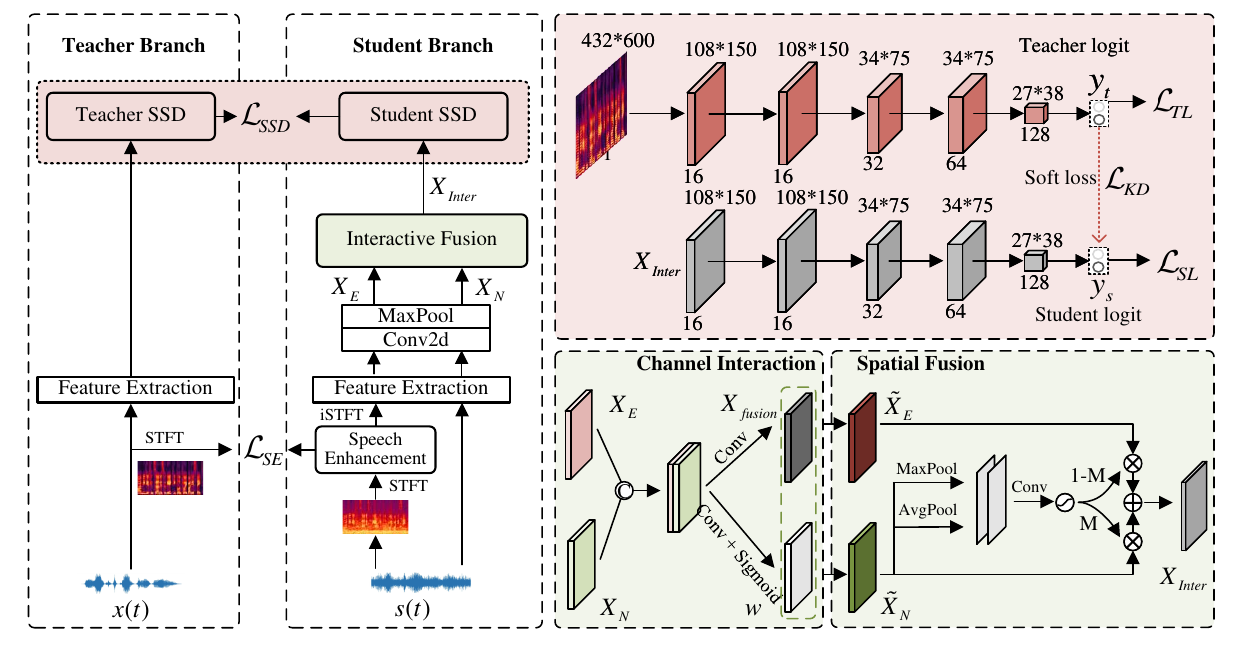}
	\caption{The structure and details of the dual-branch knowledge distillation synthetic speech detection system. Green boxes represent interactive fusion, and red boxes represent knowledge distillation. The teacher model classifier has the same structure as the student model. $C$ indicates the concatenation operation, $\otimes$ means element-wise multiplication.}
	\label{fig:model}
\end{figure*}

\section{The proposed method}
As illustrated in Fig.~\ref{fig:model},  our proposed method comprises dual branches of the teacher-student model. The teacher branch is trained using clean speech, while the student branch is trained using noisy speech. At the front end of the student branch, adaptive interactive fusion with noisy speech is conducted after speech enhancement, aiming to mitigate noise interferences, address speech distortion, and facilitate noisy speech in learning the initial data distribution of clean speech. During classification, the student’s output logit is mapped to the teacher’s logit via distillation loss, which enables the student to acquire the decision-making capacity of the teacher model. Using joint training to simultaneously optimize the entire structure, enabling the interactive fusion modules to generate features that are more beneficial to the SSD task.

\subsection{Speech Enhancement}
The purpose of speech enhancement is to remove the noise from the noisy speech and estimate the target clean speech, which can only be trained using parallel clean and noisy data. We choose some datasets as clean datasets, and add noise to them to form corresponding simulated noisy datasets, details can be found in Section \uppercase\expandafter{\romannumeral4}. A noisy speech can be expressed as:
\begin{equation}
	s(t) = x(t) + n(t)
	\label{eq1}
\end{equation}
where $s(t)$, $x(t)$ and $n(t)$ denote source noisy, clean and noise time-domain speech, respectively. We perform the short-time Fourier transform (STFT) on the source noisy speech to obtain the magnitude spectrum $S(f, t)$ as the input feature, $(f, t)$ denotes the index of frequency-time bins. And use the magnitude spectrum $X(f,t)$ corresponding to the clean speech as the training target:
\begin{align}
	&	S(f, t), \varphi(f, t) = \operatorname{STFT}(s(t)) \\
	&\widehat{X}(f, t)  = \operatorname{SE}(S(f, t), X(f,t))  \\
	&\widehat{x}(t)  = \operatorname{iSTFT}(\widehat{X}(f, t)\cdot \varphi(f, t))
	\label{eq2}
\end{align}
where $\text {SE}(\cdot)$ denotes speech enhancement network, $\widehat{X}(f, t)$ represents the output enhanced magnitude feature. After SE, $\widehat{X}(f, t)$ is multiplied with the phase spectrum $\varphi(f, t)$ of the noisy speech, and the inverse short-time Fourier transform (iSTFT) is applied to generate the enhanced speech $\widehat{x}(t)$ in the time domain. Use mean square error (MSE) as the loss function, defined as follows:
\begin{equation}
	\mathcal{L}_{SE} = \frac{1}{FT} \sum_{t=1}^{T} \sum_{f=1}^{F} |\widehat{X}(f, t) - X(f, t)|^2
	\label{eq1}
\end{equation}

After denoising, $\widehat{X}(f, t)$ is free from most noise interference. However, since MSE loss tends to emphasize the minimization of magnitude errors and ignores the details of the speech signal, it usually causes speech distortion problems. Therefore, we propose an interactive fusion module to trade off between denoising and retaining information.

\subsection{Interactive Fusion}
The purpose of speech enhancement is to restore a clean magnitude spectrum. However, due to inconsistent training targets, it is often challenging to avoid distortion issues. Directly classifying features extracted from denoised speech can adversely affect downstream tasks. We believe this is because distorted speech lacks fine structure, which can either destroy original artifacts or introduce new unknown artifacts. Some studies observe that artifacts \cite{iwamoto22_interspeech} have a particularly negative impact on downstream tasks, while the impact of noise is relatively limited. To guide the noisy branch to learn a similar data distribution as the teacher branch, we propose interactive fusion module, which fuses the original noise spectrum with the denoised spectrum. It is a trainable embedded module that is optimized together with the classifier during training.

Considering that SSD pays more attention to the details of frequency and usually requires features with a high-frequency resolution, we use a longer window length to re-extract the logarithmic amplitude spectrum features from the time domain feature $\widehat{x}(t)$ generated by the SE module. According to the more robust characteristics of the low-frequency band, the frequency region of the logarithmic magnitude spectrum (LMS) feature is directly sliced and the first half is taken to obtain the low-frequency feature:
\begin{equation}
	\widehat{X}(f,t) =log(abs(\operatorname{STFT}(\widehat{x}(t))))_{0-4KHz}  \\
	\label{eq1}
\end{equation}

To capture more spectral information and local features, and to align with the features of the teacher branch, interactive fusion is divided into two parts: channel interaction and spatial fusion. We perform $7\times7$ convolution and pooling on the LMS features extracted from the source speech $s(t)$ and $\widehat{x}(t)$ to generate multi-channel frequency-time features ${X}_{N}$ and ${X}_{E}$. Both features are expanded to 16-channel features, requiring interaction of information across all channels initially. They are concatenated in the channel dimension: $X_{concat }=\mathcal{C}\left(X_{E} , X_{N}\right)$. Generate fusion feature $X_{fusion}$ and fusion weight $w$ based on $X_{concat}$, the formula is as follows:
\begin{align}
	&X_{fusion} =conv1\left(X_{concat}\right)\\
	&w =\sigma \left(conv2\left(X_{concat}\right)\right)
\end{align}
where $conv1$ and $conv2$ denote distinct $3\times3$ convolutional layers, and $\sigma$ represents the sigmoid activation function. Thus, the enhanced feature $\widetilde{X}_{E}$ and noisy feature $\widetilde{X}_{N}$ after preliminary interaction are expressed as:
\begin{align}
	\widetilde{X}_{E} = {X}_{fusion} + w\otimes {X}_{E}\\
	\widetilde{X}_{N} = {X}_{fusion} + w\otimes {X}_{N}
\end{align}

Given our aim to preserve as much useful information as possible during denoising, in this step, we need to concentrate on the interactive feature ${X}_{fusion}$, which contains all the information. The matrix $w$ is utilized to weight the enhanced features and noisy features.

After obtaining the interactive features, spatial information fusion needs to be performed. In the context of speech features, spatial information refers to frequency-time domain features. Unlike the interaction in the previous step, this fusion step focuses more on determining the importance of each point in the frequency-time space. Noisy features contain more information, therefore, the spatial mask matrix is calculated based on the noisy features. To aggregate spatial information, average pooling is commonly employed. This technique smooths the entire feature map to derive overall features. Conversely, max pooling retains the most important features in the feature map, capturing salient features. Therefore, we utilize both pooling operations. Perform maximum pooling and average pooling on $\widetilde{X}_{N}$ separately, then concatenate the resulting feature maps in the channel dimension. Next, obtain the mask matrix $M$ after applying convolution and sigmoid activation. Finally, fuse $\widetilde{X}_{E}$ and $\widetilde{X}_{N}$ according to this mask to obtain the final interactive feature ${X}_{Inter}$:
\begin{equation}
	{X}_{Inter} = (1-M)\otimes \widetilde{X}_{E} + M\otimes\widetilde{X}_{N}
\end{equation}

The interactive fusion module adaptively combines features through channel interaction and spatial information fusion, respectively, meeting the requirements of denoising and distortion alleviation. We posit that enhanced feature serves to provide information for speech segments that would otherwise be silent, while noisy feature serves to mitigate speech distortion in segments with human voices. This assertion is further supported by the feature visualization (Fig.~\ref{fig:speech}) in the experimental section and the statistical analysis (Fig.~\ref{fig:statics}) of the mask matrix.
\begin{figure*}[!t]
	\centering
	\includegraphics[width=0.8\linewidth]{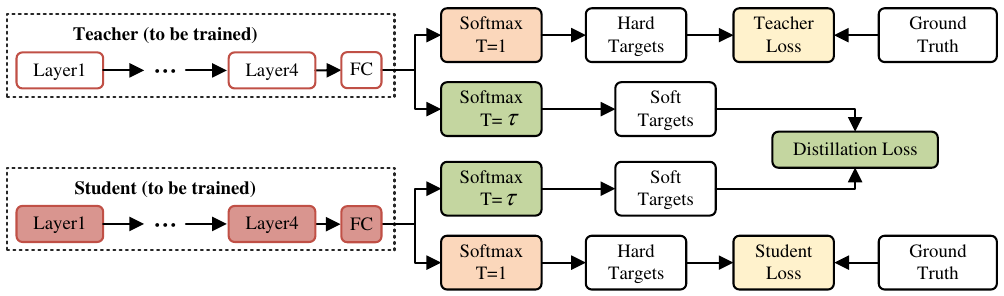}
	\caption{Schematic diagram of the structure of the teacher-student model. The teacher and student are trained in parallel, and the classification loss is calculated with the ground truth label respectively, and the loss of the soft targets between the teacher and the student is calculated at the same time. The degree of softening is controlled via the temperature hyperparameter T.}
	\label{fig:ts}
\end{figure*}

\subsection{Knowledge Distillation}
Knowledge distillation guides student branch learning from the perspective of neural responses. As illustrated in Fig.~\ref{fig:ts}, both the teacher and the student engage in online distillation, with parameters updated simultaneously. The neural response of the last output layer of the teacher model is softened, and the distillation loss is then calculated based on the soft targets of both the teacher and the student. The input data for the teacher and student models are different, and distillation at the feature level may lead to convergence difficulties. Soft logit represents a class probability distribution, thus response-based distillation enables students to directly mimic the teacher's final prediction. Since there is no need to reduce the number of model parameters and the network structures of teachers and students are the same, we use SeNet34 \cite{2021The} as the backbone. This model demonstrates excellent performance on the ASVspoof 2019 LA evaluation set. During training, the teacher model utilizes clean speech $x(t)$, while the student model employs the fused feature $X_{Inter}$ as input.

\textbf{Hard Loss.} Hard loss refers to the loss computed between model predictions and labels. Let $y_t$ and $y_s$ denote the logit of the last output layer of the teacher and student models, respectively. The loss between the ground truth label $L$ is computed using A-softmax \cite{liu2017sphereface}. The hard loss for both the student model and teacher model can be expressed as follows:
\begin{align}
	\mathcal{L}_{S L}=\operatorname{Asoftmax}\left(y_{s}, L\right) \\
	\mathcal{L}_{TL}=\operatorname{Asoftmax}\left(y_{t}, L\right)
\end{align}

\textbf{Soft Loss.} Soft loss refers to the distillation loss computed between the softened logits of the teacher and the student. The hyperparameter $\tau$ represents the distillation temperature, which regulates the degree of softening in the logits output by the teacher network. The larger $\tau$ is, the higher the degree of softening and the smoother the transition between categories. Higher temperatures lead to smoother probability distributions, facilitating the transfer of more knowledge, but they may also decrease the accuracy of the student model. The softened logit $y^{'}_{s} = y_{s}/\tau$ of the student aims to match the soft output $y^{'}_{t} = y_{t}/\tau$ of the teacher model via a Kullback–Leibler divergence (KL):
\begin{align}
	\mathcal{L}_{K D}=\tau^{2} \operatorname{KLD}\left(softmax(y^{'}_{s}), softmax(y^{'}_{t})\right)
\end{align}

Response-based knowledge enables the student to predict like the teacher. The loss of the entire synthetic speech detection network consists of both the soft loss and the hard loss, where $\alpha$ is used to control the trade-off between student loss and KD loss:
\begin{align}
	\mathcal{L}_{SSD}=(1-\alpha) \mathcal{L}_{S L}+\alpha \mathcal{L}_{KD}+ \mathcal{L}_{TL}
\end{align}

During knowledge distillation, significant disparities in ability may necessitate the intervention of a teacher assistant \cite{mirzadeh2020improved} to facilitate effective learning for students. Therefore, we employ an online distillation method, where teacher and student learn simultaneously and update parameters concurrently. The advantage of online distillation lies in the minimization of the gap between the prediction results of teachers and students, which facilitates the student model in better imitating the behavior of the teacher model. If the teacher uses pre-training, it may pose challenges for the student model to accurately mimic the teacher's predictions on data with a significant ability gap, such as low SNR.

\subsection{Joint Training}
As the front-end module of the student branch, the training objective of speech enhancement is to reconstruct a clean spectrum. If this module is pre-trained or optimized independently, it may inadvertently remove original artifacts or introduce imperceptible noise because its objective differs from the final classification task. To enable SE to estimate target speech beneficial for SSD, we integrate SE and SSD for joint training. The loss function of joint training is formulated as:
\begin{align}
	\mathcal{L}= \mathcal{L}_{SSD} + \mathcal{L}_{SE}
\end{align}

During the backpropagation process, the speech enhancement module is influenced by the synthetic speech detection network, and joint training facilitates mutual feedback between the models, leading to global optimization.
\section{Experiments}

\begin{table}[!t]
	\begin{center}
		\setlength{\abovecaptionskip}{0cm}
		\setlength{\belowcaptionskip}{0ex}
		\renewcommand\tabcolsep{8pt}
		\caption{The noisy datasets structure used by all systems except the Noise-Free baseline system.}
		\label{tab:dataset}
		\renewcommand{\arraystretch}{1.3}
		\begin{tabular}{l|l|c|c}
			\bottomrule \hline 
			\multicolumn{2}{l|}{\textbf{Subset}}& \textbf{SNRs}  & \textbf{Noise corpus}  \\
			\hline
			\multicolumn{2}{l|}{Training} & \begin{tabular}[c]{@{}c}randomly sampled \\ between [0, 20]dB \end{tabular} & 100 Nonspeech Sounds \\
			\hline
			\multicolumn{2}{l|}{Development} & \begin{tabular}[c]{@{}c}randomly sampled \\ between [0, 20]dB \end{tabular} & 100 Nonspeech Sounds \\
			\hline
			\multirow{2.5}{*}{Test} & seen  & \{0, 5, 10, 15, 20\}dB & 100 Nonspeech Sounds \\
			\cline{2-4} & unseen & \{0, 5, 10, 15, 20\}dB &  NOISEX-92\\
			\hline \toprule
		\end{tabular}
	\end{center}
\end{table}

\subsection{Datasets}
\subsubsection{Clean Datasets} The ASVspoof 2019 challenge offers a large-scale dataset comprising two subsets: LA and PA. Both subsets consist of pure human voices without noise. For this study, we selected the LA subset as the primary experimental dataset. The ASVspoof 2019 dataset continues to be a prominent resource, prompting us to conduct experiments based on it. The issue of silence \cite{muller21_asvspoof} observed in ASVspoof 2019 also persists in ASVspoof 2021, an aspect we have not yet addressed. Our experiments are focus on enhancing performance in noisy environments. In future investigations, we plan to explore performance improvements following the removal of silent segments. To investigate generalization, the ASVspoof2015 is selected for cross-dataset experiments. Therefore, the original datasets of ASVspoof 2015 and ASVspoof 2019 LA are regarded as clean datasets.

\subsubsection{Noisy Datasets} In this paper, the experimental scenario simulates noisy conditions by artificially introducing noise to the ASVspoof 2019 LA dataset. For both the training and development sets, we add random noise during training, with the SNR interval ranging from 0 to 20 dB. The noise is sourced from the 100 Nonspeech Sounds \cite{hu2010tandem} dataset \footnote{http://web.cse.ohio-state.edu/pnl/corpus/HuNonspeech/HuCorpus.html}. For the test set, we randomly select noise from the NOISEX-92 corpus \cite{varga1993assessment} and add noise to the evaluation set. The SNR is randomly chosen from the set \{0, 5, 10, 15, 20\}dB. The resulting new noisy evaluation dataset is named \textbf{unseen}. Then, we randomly select noise from the 100 Nonspeech Sounds dataset and add it to the evaluation set, named \textbf{seen}, as shown in Table~\ref{tab:dataset}. It is a widely used dataset \cite{hu2010tandem}. The specific contents of the dataset are as follows: N1-N17: Crowd noise; N18-N29: Machine noise; N30-N43: Alarm and siren; N44-N46: Traffic and car noise; N47-N55: Animal sound; N56-N69: Water sound; N70-N78: Wind; N79-N82: Bell; N83-N85: Cough; N86: Clap; N87: Snore; N88: Click; N88-N90: Laugh; N91-N92: Yawn; N93: Cry; N94: Shower; N95: Tooth brushing; N96-N97: Footsteps; N98: Door moving; N99-N100: Phone dialing. In cross-dataset experiments, we apply the same procedure to the ASVspoof 2015 dataset. The test set of ASVspoof 2021 LA already contains channel noise. Therefore, ASVspoof 2015 and ASVspoof 2019 LA produce two noisy test sets: unseen and seen. Additionally, the test set of the ASVspoof 2021 LA dataset itself serves as a noisy dataset. Table~\ref{tab:noisy} illustrates the composition of the noisy datasets used by all systems (except Noise-Free).

\subsubsection{Experimental Datasets} In the experiment, we initially evaluate the ASVspoof 2019 LA \textbf{clean} set, the \textbf{unseen} noisy dataset, and the \textbf{seen} noisy dataset. Subsequently, we conduct offline speech enhancement on both the \textbf{unseen} and \textbf{clean} sets of ASVspoof 2019 LA and test the models ``Noise-Free", ``MCT1" and ``MCT2" on the above datasets. Next, we conduct cross-dataset experiments on the original \textbf{clean} test set, the \textbf{unseen} noisy dataset, and the \textbf{seen} noisy dataset of ASVspoof 2015, respectively. Additionally, cross-dataset experiments are also conducted on the ASVspoof 2021 LA test set. For the ablation experiment, we select the \textbf{unseen} dataset of ASVspoof 2019 LA as the set.

\subsection{Experimental Setup}
\subsubsection{Baselines} In this work, to study the effectiveness of the proposed method, we choose SENet-34 \cite{2021The} as the backbone SSD network for all baseline models. The speech enhancement module uses a convolutional recurrent neural (CRN) model \cite{tan2018convolutional}. Details of the different baselines used for comparison in this work can be found in Table~\ref{tab:models}. The first three models are traditional structures, consistent with Fig.~\ref{fig:structure}(a). The structures of ``Cascade" and ``Joint" are consistent with Fig.~\ref{fig:structure}(b). The ``Cascade" system simply concatenates the speech enhancement model and the synthetic speech detection model. During training, the loss of the speech enhancement part is backpropagated independently, resulting in less correlation between the frontend and backend. Conversely, the ``Joint" system jointly trains the speech enhancement model and the synthetic speech detection model. During actual training, both tasks need to be accomplished, and backpropagation is performed simultaneously. This approach affects the learning of shared parameters and is beneficial for optimizing the speech enhancement part towards the final objective. Compared to the ``Joint" system, the structure of ``DKDSSD" differs in that it adopts a dual-branch knowledge distillation structure and the interactive fusion module.

\begin{table}[!t]
	\setlength{\abovecaptionskip}{0cm}
	\setlength{\belowcaptionskip}{-3ex}
	\caption{The configuration of different systems in this work. ``$\dagger$" means proposed.}
	\label{tab:models}
	\centering
	\begin{tabularx}{\linewidth}{lX}
		\hline \toprule
		\textbf{Systems name}&  \textbf{Details} \\
		\midrule Noise-Free \cite{2021The}  & System trained with only clean data. \\ 
		\midrule MCT1 \cite{qian2017deep}  & System trained with half noisy speech and half clean. \\
		\midrule MCT2 \cite{qian2017deep}  & System trained with completely noisy speech. \\
		\midrule Cascade \cite{subramanian2019speech}  & Cascaded system of SSD and SE, trained with noisy speech. The system is optimized using SSD training objective only.\\
		\midrule Joint \cite{ma2021multitask}  & Multitask-based joint training system of SSD and SE, trained with noisy speech. \\
		\midrule DKDSSD $\dagger$ & Proposed dual-branch system, trained with pairs of clean and noisy speech. \\
		\hline\toprule
	\end{tabularx}%
\end{table}%

\subsubsection{Implementation Details} For SE, we use 161-dimensional magnitude spectrums as input features, and the time frames are set to 600. For SSD, to extract log magnitude spectrogram features, we set the blackman window length and hop length of STFT to 1728 and 130,  respectively. The speech is truncated or spliced so that the number of frames of all input features remains the same as 600 frames. Due to the study in \cite{2021The,xue2022audio}, the low sub-band log magnitude spectrogram feature is used in our work, so all the input features of synthetic speech detection maintain the same shape of 433 × 600. Additionally, Adam is the optimizer of both SE and SSD, with the hyperparameter $\alpha$ set to 0.05 and the temperature $\tau$ set to 3. 
These hyperparameters are determined after several experiments. We train each system for 32 epochs, and the model with the lowest loss on the development set is selected as the final model for evaluation. For the DKDSSD model, the teacher model and the student model are trained in parallel during training, and only the student model is used for inference. We evaluate the performance of all systems using the equal error rate (EER), a metric that reflects the ability to detect synthetic speech. Additionally, the min tandem detection cost function (t-DCF) is used as the evaluation metric for ASVspoof 2021, which is officially considered a more important metric on this dataset.

\begin{table*}[!t]
	\renewcommand\tabcolsep{8pt}
	\caption{Results in terms of EER(\%) for \textbf{ASVspoof 2019 seen}, \textbf{unseen} and \textbf{clean} set. ``AVG." is not the average of the EERs for the 5 SNR cases, but a result of the total evaluation set. The EER of each SNR is calculated separately. ``$\dagger$" means proposed.}
	\label{tab:noisy}
	\centering
	\begin{tabular}{lcccccc|cccccc|c}
		\hline\toprule
		\multirow{2}{*}{Systems}&\multicolumn{6}{c|}{\textbf{seen}} & \multicolumn{6}{c|}{\textbf{unseen}} & \multicolumn{1}{c}{\textbf{clean}} \\
		\cmidrule{2-14}& 0dB	& 5dB	& 10dB	& 15dB	& 20dB & AVG. & 0dB	& 5dB	& 10dB	& 15dB	& 20dB & AVG. & AVG.\\
		\midrule
		Noise-Free \cite{2021The}  	& 44.91 & 39.70	&26.30 & 19.90 &18.46 & 27.98&55.53 &53.40 &34.12 &21.81 &19.36 &36.04 & \textbf{2.21} \\
		\midrule
		MCT1 \cite{qian2017deep}	&  7.91 & 5.16  & 4.14 & 3.83 &  2.99  &  5.01  & 14.04 & 11.02 & 7.46 & 5.69 & 3.76 & 8.59 & 2.43\\
		\midrule
		MCT2 \cite{qian2017deep}                 	 	&  6.72  & 4.39&3.52  &3.42 &2.86&4.27&12.03&9.10&5.57&4.37&3.08&7.16 & 3.37\\
		\midrule
		Cascade \cite{subramanian2019speech}      & 7.53 &6.13& 5.04& 4.45 &4.57 &5.74&  12.10 & 8.49	&6.22  & 5.27&4.10&7.60 &4.01\\
		\midrule
		Joint \cite{ma2021multitask}         		&  6.80 & 4.89	&3.73  & 3.89&3.52&4.74&10.33&7.94&5.50&4.51&3.41&6.92 &3.29\\
		\midrule
		DKDSSD $\dagger$       &  \textbf{5.26} & \textbf{3.82}	&\textbf{2.83}  &\textbf{2.93} &\textbf{2.39}&\textbf{3.55}&\textbf{8.52}&\textbf{6.53}&\textbf{4.58}&\textbf{3.41}&\textbf{2.95}&\textbf{5.40} & 3.28\\
		\bottomrule\hline
	\end{tabular}
\end{table*}

\begin{table}[!t]
	\renewcommand\tabcolsep{4pt}
	\renewcommand\arraystretch{1.2}
	\caption{Results in terms of EER(\%) for \textbf{ASVspoof 2019 unseen} set and \textbf{clean} set. It represents the result of \textbf{performing SE} on the set first and then testing, reflecting the benefits and limitations of SE.}
	\label{tab:distortion}
	\centering
	\begin{tabular}{lcccccc|c}
		\hline\toprule
		\multirow{2}{*}{Systems}&\multicolumn{6}{c|}{\textbf{unseen}} & \multicolumn{1}{c}{\textbf{clean}}  \\
		\cmidrule{2-8}& 0dB	& 5dB	& 10dB	& 15dB	& 20dB & AVG. & AVG. \\
		\midrule
		Noise-Free \cite{2021The}     	  &37.11&30.27&19.83&18.68&18.19&23.48 &15.66 \\
		\midrule
		MCT1 \cite{qian2017deep}  	&21.23&18.03&15.71&13.67&11.92&16.14 & 9.00 \\
		\midrule
		MCT2 \cite{qian2017deep}        	&29.83&25.73&20.56&15.34&11.01&20.92 & 8.67\\
		\bottomrule\hline
	\end{tabular}
\end{table}

\begin{table*}[htbp]
	\centering
	\renewcommand\tabcolsep{2.2pt}
	\renewcommand{\arraystretch}{1.2}
	\caption{Results in EER(\%) for all baseline models on the \textbf{ASVspoof 2019 unseen} set. The 5 types of SNR conditions under all 12 noisy scenarios are counted separately. The unseen data set is randomly generated by adding noise with different SNRs. The number of real and fake audio for each type of noise is basically the same.}
	\label{tab:noisefig}
	\begin{tabular}{|c|c|c|c|c|c|c|c|c|c|c|c|c|c|c|c|}
		\hline \textbf{Noise}  & \textbf{SNR}  & \textbf{Noise-Free} & \textbf{MCT1} & \textbf{MCT2} & \textbf{Cascade}  & \textbf{Joint} & \textbf{DKDSSD} &  \textbf{Noise}  & \textbf{SNR}  & \textbf{Noise-Free} & \textbf{MCT1} & \textbf{MCT2} & \textbf{Cascade}  & \textbf{Joint} & \textbf{DKDSSD}\\
		\hline \multirow{6}{*}{white} & 0 &  67.42&  9.10 &  6.42 &  8.32 &  7.40 &  \textbf{6.37} & \multirow{6}{*}{ m109} &  0 &  68.96 &  9.23 &  10.80 &  10.28 &  9.39 &  \textbf{6.88} \\
		\cline{2-8} \cline{10-16}
		& 5 & 66.79 & 7.61& 5.93& 6.08&6.94 &\textbf{4.91} & & 5 & 32.64&8.91 &6.75 &7.12 & 7.94&\textbf{6.11} \\
		\cline{2-8} \cline{10-16}
		& 10 & 53.04 &4.35 & 2.62& 5.24& 3.46& \textbf{2.62}& & 10 & 20.00& 5.74& 5.04&5.74 &4.33 &\textbf{4.18} \\
		\cline{2-8} \cline{10-16}
		& 15 & 21.48 &2.79 & 2.94& 5.42&3.72 &\textbf{2.79} & & 15 & 18.47 &5.05 & 3.38& 5.05&3.43 &\textbf{2.50} \\
		\cline{2-8} \cline{10-16}
		& 20 & 19.73 &3.56 & 3.76& 4.53& 3.76& \textbf{2.03}& & 20 & 18.67 &2.83 &\textbf{1.85} & 2.83& 3.75&2.83 \\ 
		\cline{2-8} \cline{10-16}
		& AVG. & 21.48 &5.74 &4.47 &5.81 &4.88 & \textbf{3.61}& & AVG. & 30.58 &6.49 &5.64 &6.71 &5.64 &\textbf{4.79} \\ \hline 
		\hline \multirow{6}{*}{f16} & 0 & 72.95 &  11.64 &  8.81 & 10.36 &  9.07 &  \textbf{7.93} & \multirow{6}{*}{ hfchannel} &  0 &  67.76 &  6.59 &  5.66 &  7.53 &  6.70 &  \textbf{4.78} \\
		\cline{2-8} \cline{10-16}	& 5 & 69.21 &9.56 &9.51 & 7.46& \textbf{6.20}& 6.36& & 5 & 66.11& 4.18& 3.29& 4.13& 5.75& \textbf{2.46}\\
		\cline{2-8} \cline{10-16}
		& 10 & 32.50  & 5.94& 3.62& 5.99&4.30 &\textbf{3.48} & & 10 & 37.36& 1.76&1.76 & 3.56&2.59 & \textbf{1.66}\\
		\cline{2-8} \cline{10-16}
		& 15 & 17.69 & 6.34& 3.34& 3.56& 3.62&\textbf{3.34} & & 15 & 17.20& 3.65& 4.57& 5.48& 5.48&\textbf{2.89} \\
		\cline{2-8} \cline{10-16}
		& 20 & 19.04  & 4.15&\textbf{3.02} & 3.94& 4.04& 3.07& & 20 & 15.64& 2.63&1.75 &2.63 & 2.68& \textbf{1.75}\\ 
		\cline{2-8} \cline{10-16}
		& AVG. & 41.08 &8.23 & 6.29& 6.46&5.88 &\textbf{5.13} & & AVG. & 40.67 &3.58 &3.19 & 4.76& 4.60& \textbf{3.01}\\
		\hline
		\hline \multirow{6}{*}{machinegun} & 0 &  19.80 &8.72 &7.92 &13.89 & 9.93 &  \textbf{7.02} & \multirow{6}{*}{ buccaneer1} & 0 & 71.48 &  15.22 &  12.41 &  13.35 & 11.21 & \textbf{9.44} \\
		\cline{2-8} \cline{10-16}
		& 5 &19.78 & 7.14& 3.70& 8.94& 5.29& \textbf{3.40}& & 5 & 75.24&10.63 &7.14 &7.87 & 8.01& \textbf{6.02}\\
		\cline{2-8} \cline{10-16}
		& 10 & 18.46&5.16 & 4.30& 6.01& 5.06& \textbf{3.34}& & 10 & 58.00&7.43 &5.77 & 6.86&5.77 &\textbf{4.31} \\
		\cline{2-8} \cline{10-16}
		& 15 & 19.03&5.78 &3.31 &6.82 & 5.82& \textbf{3.27}& & 15 &23.97 &5.34 &4.48 &2.82 &3.58 & \textbf{2.57}\\
		\cline{2-8} \cline{10-16}
		& 20 & 16.94& 4.30&\textbf{2.59} &3.47 &3.42 &3.27 & & 20 & 16.68& \textbf{2.52}& 3.37& 3.47& 3.37&3.31 \\ 
		\cline{2-8} \cline{10-16}
		& AVG. & 19.48	& 6.19&4.24 &9.48 &6.66 &\textbf{3.86} & & AVG. & 44.90 & 9.16& 7.41& 8.09& 7.04& \textbf{5.07}\\
		\hline
		\hline \multirow{6}{*}{factory2} & 0 &  68.95 & 13.01 & 8.04& 9.02 & 9.02& \textbf{6.86} &   \multirow{6}{*}{ factory1} &  0 &  68.78 &  13.54 &  \textbf{9.67} & 11.38&  10.52 & 9.72 \\
		\cline{2-8} \cline{10-16}
		& 5 & 45.47& 11.09& 7.29&7.90& 7.95& \textbf{6.18}& &  5 & 71.31& 11.06& 7.92& 8.73&8.89 & \textbf{7.70}\\
		\cline{2-8} \cline{10-16}
		& 10 &20.09 &7.89 &5.27& 7.03&6.13 & \textbf{4.37}& &  10 &43.26 &7.14 & 4.29& 5.39&4.49 &\textbf{3.59} \\
		\cline{2-8} \cline{10-16}
		& 15 &20.70 &5.06 & 5.96&5.96& 5.16&\textbf{3.97} & &  15 &21.26 &5.29 &3.49 &3.54 & 3.49&\textbf{2.51} \\
		\cline{2-8} \cline{10-16}
		& 20 & 21.18&3.81 &3.45&3.86 & \textbf{3.45}&3.61 & &  20 &20.81 &4.11 & 3.28& 4.79&3.28 & \textbf{3.23}\\ 
		\cline{2-8} \cline{10-16}
		& AVG. & 34.41 & 5.93&8.62& 7.03& 6.51&\textbf{5.35} &  & AVG. & 43.36 & 8.37& 5.95& 7.21& 6.64& \textbf{5.60}\\
		\hline
		\hline \multirow{6}{*}{destroyerops} & 0 &  76.19 & 14.18&11.69 & 10.93 & 9.69& \textbf{8.64} & \multirow{6}{*}{ pink} &  0 &  70.80 & 15.11 & 10.62&12.28 & 11.48 & \textbf{10.57}\\
		\cline{2-8} \cline{10-16}
		& 5 & 65.44&10.94 &10.84 &9.20 & 8.60& \textbf{7.01} & & 5 &75.65 &12.58 &8.80 & \textbf{7.09}& 8.30& 8.00\\
		\cline{2-8} \cline{10-16}
		& 10 &27.38 &7.36 &6.33 &5.30 & 6.43&\textbf{5.25} & & 10 &55.45 & 7.28& 6.19& 5.58&6.35 &\textbf{5.37} \\
		\cline{2-8} \cline{10-16}
		& 15 & 17.25&4.38 &3.60 & 4.22& 3.60&\textbf{2.11} & & 15 &21.81 &5.21 & 5.21& 5.21& 4.22&\textbf{5.21} \\
		\cline{2-8} \cline{10-16}
		& 20 &19.97 &4.33 & \textbf{1.92}& 3.49& 3.44&3.44 & & 20 &19.84 & 4.87&2.51 &4.15 & 3.18& \textbf{2.31}\\ 
		\cline{2-8} \cline{10-16}
		& AVG. & 46.40&8.64 & 8.45& 7.68& 7.01&\textbf{5.60} & & AVG. & 44.55 &9.07 &6.76 &7.10 & 6.91& \textbf{6.32}\\
		\hline
		\hline \multirow{6}{*}{babble} & 0 &  77.86 &16.89 &16.28 & 15.38 & 14.63 & \textbf{12.08} & \multirow{6}{*}{ leopard} &  0 &  18.31 &15.59 & 13.78& 12.78 & 9.16 & \textbf{9.11} \\
		\cline{2-8} \cline{10-16}
		& 5 &59.40 &13.30 &12.40 & 11.55& 11.05&\textbf{8.55} & & 5 & 17.91&13.03 & 11.18& 7.95& 8.11& \textbf{6.62}\\
		\cline{2-8} \cline{10-16}
		& 10 &26.95 &10.71 & 6.25& 5.38&6.04 & \textbf{5.38}& & 10 & 19.96 &12.60 & 8.20& 8.00&8.65 & \textbf{7.02}\\
		\cline{2-8} \cline{10-16}
		& 15 & 18.20&7.72 &6.55 & 6.86& 6.76& \textbf{5.54}& & 15 & 18.51 &7.70 &\textbf{3.90} & 6.04& 5.85& 4.87\\
		\cline{2-8} \cline{10-16}
		& 20 &17.77 &3.94 &3.00 &3.73 & 3.83&\textbf{2.80} & & 20 &20.83  &4.97 &4.97 &6.42 &\textbf{3.47} & 4.97\\ 
		\cline{2-8} \cline{10-16}
		& AVG. & 44.29	&11.15 &11.52 &10.25 &9.75 &\textbf{6.97} & & AVG. & 19.13 &11.66 &9.12 &8.28 & 7.55&\textbf{6.54} \\
		\hline
	\end{tabular}
\end{table*}

\section{Experimental Results}
\subsection{Results on ASVspoof 2019 LA}
\subsubsection{Effectiveness and Limitation of SE}
We first analyze the effectiveness and limitations of SE. From Table~\ref{tab:distortion}, we observe that the SE process significantly improves the performance of the ``Noise-Free" in noisy conditions, while the performance of ``MCT1" and ``MCT2" degrades. For the ``Noise-Free" trained only on clean speech, noise poses the biggest distraction, and denoising leads to performance improvement. However, for MCT systems, distortion proves to be more disruptive than noise. On the clean set, denoising degrades the performance of ``MCT1", ``MCT2" and ``Noise-Free" significantly. This is because for clean data, SE is unnecessary. while SE is beneficial for reconstructing speech from noisy conditions, processing distortion often introduces unseen artifacts. Therefore, it becomes essential to address the distortion problem alongside denoising.

\subsubsection{Comparison with Other Baselines}
Table~\ref{tab:noisy} shows the results of all SSD systems. We can observe that models trained only in clean conditions lack noise robustness, especially in the case of low SNR. Therefore, it is necessary to improve the generalization of SSD systems to deal with more complex real-world situations. The overall performance of the ``MCT1" model is much better than that of the ``Noise-Free", and it is more stable when the noise is strong because it is trained with both clean and noisy speech together. Multi-condition training allows neural networks to model more discriminative feature representations, thereby improving detection ability under noisy conditions. However, due to the overfitting of noise, the performance of multi-condition training on the clean data set is degraded, and the degradation of ``MCT2", trained with only noisy data, is more obvious.

``Cascade" uses speech enhancement as the front end, which has a significant performance improvement compared to the system without SE, but it is not as good as the performance of ``MCT2". This is because the cascade method only optimizes the synthetic speech detection and does not optimize the speech enhancement. When joint training is used for multi-task learning, the ``Joint" system is further improved on the basis of the cascaded system, which shows that joint training is beneficial to promote the global optimization of the overall structure. In clean conditions, it also performs well. The reason is that the models can share information and influence each other.

All systems degrade significantly on unseen noisy datasets, indicating that unseen noise remains a difficult problem to deal with. Our proposed ``DKDSSD" system achieves the lowest EER of 3.55\% and 5.40\% on the seen and unseen datasets, respectively, and performs best in all SNR situations, indicating that our proposed method can effectively improve the detection ability in noisy scenes and has better generalization in unseen noisy situations.

\subsubsection{Result For All Unseen Noisy Conditions}
Table~\ref{tab:noisefig} shows the EER of the five models in all noisy scenarios. It can be observed that DKDSSD performs best in almost all types of noise when the SNR is lower than 20dB, indicating that ``DKDSSD" has strong noise robustness. When the SNR is 20dB, the performance of ``MCT2" is second only to ``DKDSSD", because ``MCT2" also models the noise during training, and maintains a certain perception ability for unseen noise. However, when the SNR is low, the performance is not as good as the model with a speech enhancement module. Among them, babble is the most difficult noise to detect, because babble contains ambiguous human voice. It is in the same frequency band as the speaker's fundamental frequency, which will pollute the original spectrum, and it is difficult for speech enhancement methods to reconstruct speech. The overall performance of ``DKDSSD" is balanced, but it performs poorly under non-stationary noise with poor periodicities such as babble and leopard, and the noise with a relatively uniform frequency spectrum such as white noise is easier to remove.

\subsection{Cross-Dataset Results}
To demonstrate the generalization of our proposed model, we conduct cross-dataset tests on two datasets: ASVspoof 2021 LA test set and ASVspoof 2015 test set. It means that the training and development datasets are both ASVspoof 2019 LA, and only tested on two out-of-domain datasets. Among them, ASVspoof 2021 LA is the official dataset, since this dataset itself has been compressed and encoded with some telephone channel noise, we have not made any changes. We performed noise simulation on the data of the ASVspoof 2015 test. The experimental results are divided into three categories: seen, unseen, and clean. The types of seen and unseen noise and the SNR are consistent with the ASVspoof 2019 test dataset in Table~\ref{tab:dataset}. Clean means the test dataset without any processing.

\subsubsection{Results on ASVspoof 2015}
Table~\ref{tab:2015} shows the results of the ASVspoof 2015 test dataset. The ``Noise-Free" model performs poorly under all three conditions, which shows that the distribution of the dataset in 2015 is very different from that in 2019. Even under clean conditions without noise, ``Noise-Free" can only achieve an EER of 33.96\%. However, the ``DKDSSD" model still shows the best generalization ability and achieves the lowest EER in the three three types scenarios. We can see that all models perform better on seen datasets because the additive noise of seen datasets is known at training time. When faced with unseen noise, the model experiences performance degradation. This is because the speech enhancement model struggles to handle unseen noise, thus introducing unseen artifacts, consequently impacting SSD tasks.

\subsubsection{Results on ASVspoof 2021 LA}
Table~\ref{tab:2021} shows the results of the ASVspoof 2021 LA test dataset. In addition to the 5 baselines, two official best baselines (B03, B04) and two new models are also listed. Among them, ``AASIST" is a SOTA single model on ASVspoof2019. We used the open-source code of ``AASIST" to conduct cross-dataset experiments on ASVspoof 2021 LA test sets. ``Noise-Free" has the worst generalization because this test set takes into account phone encoding and transmission, which is completely unseen during training. ``MCT1" and ``MCT2" perform better, because these two models add noise during the training process, which is equivalent to doing data augmentation so that they have a perception of phone noise. From the results, it can be observed that the baseline with speech enhancement has better generalization, and ``DKDSSD" achieved the lowest EER of 4.35\%. It shows that the proposed model can not only cope with general additive noise but also maintain generalization to telephone transmission channel noise.

\begin{table*}[!t]
	\renewcommand\tabcolsep{8pt}
	\caption{Results in terms of EER(\%) for \textbf{ASVspoof 2015 seen}, \textbf{unseen} and \textbf{clean} set. ``AVG." is not the average of the EERs for the 5 SNR cases, but a result of total evaluate set. The EER of each SNR is calculated separately. ``$\dagger$" means proposed.}
	\label{tab:2015}
	\centering
	\begin{tabular}{lcccccc|cccccc|c}
		\hline\toprule
		\multirow{2}{*}{Systems}&\multicolumn{6}{c|}{\textbf{seen}} & \multicolumn{6}{c|}{\textbf{unseen}} & \multicolumn{1}{c}{\textbf{clean}} \\
		\cmidrule{2-14}& 0dB	& 5dB	& 10dB	& 15dB	& 20dB & AVG. & 0dB	& 5dB	& 10dB	& 15dB	& 20dB & AVG. & AVG.\\
		\midrule
		Noise-Free \cite{2021The}      		& 46.19 &	45.94 &	44.92 &	44.61 &	41.30 &	45.32 &	46.55 	& 49.14 &	48.94 &	47.01 &	41.40 &	46.74 &	33.96 
		\\
		\midrule
		MCT1 \cite{qian2017deep}  	&  18.39 &	15.05 &	12.42 &	11.58 &	10.97 &	14.45 &	25.07 &	19.98 &	15.31 &	13.43 &	11.81 &	17.64 &	9.33 
		\\
		\midrule
		MCT2 \cite{qian2017deep}                  	 	&  16.68 &	13.43 &	11.17 &	9.59 &	9.16 &	12.64 &	24.31 &	17.07 &	13.09 &	11.14 &	10.20 &	16.04 &	7.85 
		\\
		\midrule
		Cascade \cite{subramanian2019speech}      & 15.82 &	11.15 &	9.70 &	8.81 &	8.94 &	11.40 &	21.32 &	14.75 &	11.76 &	10.22 &	9.73 &	14.34 &	8.50 
		\\
		\midrule
		Joint \cite{ma2021multitask}         		& 15.51 &	11.71 &	10.03 &	9.20 &	9.43 &	11.81 &	20.08 &	15.14 &	12.40 &	10.77 &	9.34 &	14.59 	& 8.92
		\\
		\midrule
		DKDSSD $\dagger$       &  \textbf{14.10} &	\textbf{10.57} &	\textbf{8.88} &	\textbf{8.03} &	\textbf{7.71} &	\textbf{10.53} &	\textbf{19.92} &	\textbf{14.30} &	\textbf{11.34} &	\textbf{9.82} &	\textbf{8.27} &	\textbf{13.61} &	\textbf{7.76} 
		\\
		\bottomrule\hline
	\end{tabular}
\end{table*}

\begin{table}[!t]
	\renewcommand\tabcolsep{10pt}
	\caption{Results in terms of EER(\%) for \textbf{ASVspoof 2021 LA} dataset. B03 and B04 are the official two best baselines, AASIST, BTS-E, and SE-Rawformer are famous models with excellent performance on the 2019 and 2021 datasets.}
	\label{tab:2021}
	\centering
	\begin{tabular}{lcc}
		\hline\toprule
		Systems &  EER	& min t-DCF \\
		\midrule
		AASIST \cite{jung2022aasist}   & 10.51
		&   0.4884
		\\		
		LFCC-LCNN (B03)  \cite{yamagishi2021asvspoof} & 9.26 & 0.3445 \\ 
		RawNet2 (B04)  \cite{yamagishi2021asvspoof} & 9.50 & 0.4257 \\ 
		BTS-E \cite{doan2023bts} & 8.75 & 0.3893 \\ 
		SE-Rawformer \cite{liu2023leveraging}  & \textbf{4.53}  & \textbf{0.3088}  \\ \midrule
		Noise-Free \cite{2021The}   & 15.33
		&   0.3655 \\ 
		MCT1 \cite{qian2017deep} & 5.93 & 0.3157
		\\
		MCT2 \cite{qian2017deep}    & 5.8 &   0.3304   
		\\ 
		Cascade \cite{subramanian2019speech} & 5.28 & 0.3072 \\
		Joint \cite{ma2021multitask}  & 4.82 & 0.2969
		\\ 
		DKDSSD $\dagger$  &  \textbf{4.35} &   \textbf{0.2839}
		\\ 
		\bottomrule\hline
	\end{tabular}
\end{table}

\begin{table}[!t]
	\renewcommand\tabcolsep{4pt}
	\caption{The EER(\%) results of the ablation study on the model itself, conducted on the \textbf{ASVspoof 2019 unseen} dataset. "w/o" indicates without. IF is the interaction fusion module, KD is the knowledge distillation framework.}
	\label{tab:ablation}
	\centering
	\begin{tabular}{lccccccc}
		\hline\toprule
		\multirow{2}{*}{Systems} & \multicolumn{7}{c}{Unseen}  \\		\cmidrule{2-8}& 0dB	& 5dB	& 10dB	& 15dB	& 20dB & AVG. & params\\
		\midrule
		DKDSSD  $\dagger$    &\textbf{8.52}&\textbf{6.53}&\textbf{4.58}&\textbf{3.41}&\textbf{2.95}&\textbf{5.40} & 21.06 M\\ \midrule
		w/o joint training      &11.48&8.70&5.82&4.44&3.56&7.18& 21.06 M\\ \midrule
		w/o IF      &10.52&7.48&5.36&4.51&3.56& 6.87&20.26M\\ \midrule
		w/o KD       &12.04&8.85&5.63&4.23&3.33&7.57&19.72M\\ 
		\bottomrule\hline
	\end{tabular}
\end{table}

\subsection{Ablation Study}
\subsubsection{The Ablation Study Results of The Model Itself}
Table~\ref{tab:ablation} presents the results of the ablation experiments. We mainly analyze the results of the unseen dataset, because the unseen noise is more challenging. When joint training is not used, speech enhancement and synthetic speech detection become independent tasks, which greatly affects the optimization of the two tasks and may lead to falling into a local optimum, thus resulting in an increase in EER from 5.40\% to 7.18\%. Joint training facilitates the common optimization of all models in the system. When there are multiple modules that need to be optimized, joint training is an effective strategy.

The teacher model can guide the student branch to learn a distribution similar to clean data, and when the teacher branch is lost (w/o KD), the remaining noisy branch cannot generate noise-robust features due to the lack of clean teacher supervision. The interactive fusion module cannot adaptively fuse noisy speech without the constraint of distillation loss, and may introduce unseen artifacts. This resulted in the worst EER performance (7.57\%) in the Table~\ref{tab:ablation}.

The role of the interactive fusion module is to adaptively fuse the denoising spectrum and the noisy spectrum, which can absorb the beneficial information of the noisy spectrum and alleviate speech distortion. When there is no interactive fusion, the system directly feeds the enhanced spectrum into the back-end classifier for identification. The biggest disturbance to the classification task is the unseen artifact caused by the distortion, which affects the data distribution of the data.

\subsubsection{The Ablation Study Results of Replacing Modules} 
Table~\ref{tab:newablation} shows the ablation experimental results on the ASVspoof 2019 unseen set after replacing the module of DKDSSD. For the convenience of demonstrating the performance of each module, ``DKDSSD" is denoted as ``CRN+IF+KD+SeNet", where CRN represents the front-end speech enhancement model, IF is the interaction fusion module, KD is the knowledge distillation framework, and SeNet is the back-end spoofing detection classifier. Observation adding (OA) \cite{iwamoto22_interspeech} is a confirmed simple and effective fusion method, which reduces the impact of distortion by superimposing the original noisy speech in proportion. TSSD \cite{hua2021towards} is a network that performs well in modeling time-domain features on ASVspoof2019. It uses raw waveforms as features to avoid information loss. Diffusion \cite{richter2023speech} is a diffusion-based generative model, and using pre-trained models from Diffusion is tested to see if mature pre-trained models are more advantageous for downstream tasks.

\begin{table}[!t]
	\renewcommand\tabcolsep{1.7pt}
	\caption{The EER(\%) results of the ablation study on replacing modules, conducted on the \textbf{ASVspoof 2019 unseen} dataset. ``$\dagger$" represents proposed DKDSSD}
	\label{tab:newablation}
	\centering
	\begin{tabular}{lccccccc}
		\hline\toprule
		\multirow{2}{*}{Systems} & \multicolumn{7}{c}{Unseen}  \\		\cmidrule{2-8}& 0dB	& 5dB	& 10dB	& 15dB	& 20dB & AVG. & params\\
		\midrule
		CRN+IF+KD+SeNet  $\dagger$ &\textbf{8.52}&6.53&\textbf{4.58}&\textbf{3.41}&\textbf{2.95}&\textbf{5.40} & 21.06 M\\ \midrule
		CRN+IF+PKD+SeNet  &10.03&7.07&4.83&3.75&3.15&5.95& 21.06 M\\ \midrule
		CRN+OA\cite{iwamoto22_interspeech}+KD+SeNet      &10.11&7.14&4.84&3.94&2.86&6.37&20.26M\\ \midrule
		CRN+KD+TSSD\cite{hua2021towards}      &11.33&\textbf{6.19}&5.36&5.06&4.43&6.53&18.28M\\ \midrule
		CRN+TSSD\cite{hua2021towards}     &16.84&10.46&6.55&5.41&5.25& 10.12&17.93M\\ \midrule
		Diffuision\cite{richter2023speech}+SeNet      &20.84&19.87&20.88&20.83&21.45&21.75& 66.93M\\\midrule
		Diffuision\cite{richter2023speech}+IF+SeNet  &13.21&10.68&7.99&6.87&4.86&8.62& 67.73M\\
		\bottomrule\hline
	\end{tabular}
\end{table}

\textbf{Impact of Knowledge Distillation Methods.} The knowledge distillation (KD) method we employ involves training the teacher model and the student model together, known as online distillation. PKD denotes pre-training the teacher model first, referred to as offline distillation, involving a two-stage training process. Experimentally, after pre-training the teacher model separately (``CRN+IF+PKD+SeNet"), the results are not as good as our proposed ``DKDSSD" (``CRN+IF+KD+SeNet"). We speculate that this is because there are significant differences in predictions between the pre-trained teacher model and the student model (due to different training data), hindering the learning process of the student. Simultaneously training the teacher and student models helps to narrow this gap. Offline distillation typically requires large-scale teacher models to guide the students, and there is always a gap in capabilities between large teachers and small students, with students often heavily relying on the teacher. Our teacher-student network has the same structure and a small parameter count, making it more suitable for online distillation. In online distillation, teachers and students learn together, and the gap in their abilities remains small throughout training.

\textbf{Impact of Interaction Fusion.} To demonstrate the necessity of fusing noisy speech, we replaced the interactive fusion method with the OA method \cite{iwamoto22_interspeech}. OA is a confirmed simple and effective fusion method that reduces the influence of distortion by linearly superimposing enhanced speech and original noisy speech in proportion. Specifically, we linearly combine enhanced speech and original noisy speech in a ratio of 0.7:0.3. This hyperparameter setting is based on the optimal experimental results reported in \cite{iwamoto22_interspeech}. It can be observed that although there is a slight performance decrease compared to the methods using interactive fusion, OA still achieves good performance, indicating that the original noisy speech contains information conducive to detection. Our proposed interactive fusion method can adaptively learn the optimal fusion mask matrix through multiple training epochs, thus performing better.

\textbf{Impact of Classifier.} Experiments ``CRN+KD+TSSD" and ``CRN+TSSD" in Table~\ref{tab:newablation} are conducted for end-to-end models. TSSD, which performs well in modeling time-domain features on ASVspoof2019, is an end-to-end model. It uses raw waveforms as features to avoid information loss. From the experimental results, the performance of TSSD is not as good as our proposed ``DKDSSD" (``CRN+IF+KD+SeNet"). The drawback of manually extracting features lies in the loss of phase information, as the original waveform contains both phase and magnitude spectrum information. Although DKDSSD employs manually extracted features that lose phase information, the interaction fusion module (IF) effectively compensates for this loss. In our previous work, we explored how to fully utilize phase features and found that subband fusion methods based on complex spectra perform excellently. However, we have not yet studied complex spectral features in terms of noise resistance, which will be our future work. Additionally, TSSD with knowledge distillation (KD) shows relatively better performance, indicating that the training framework of knowledge distillation is useful.

\begin{figure}[!t]
	\centering
	\setlength{\abovecaptionskip}{0.1cm}
	\setlength{\belowcaptionskip}{-2ex}
	\subfloat[DKDSSD]{\includegraphics[width=0.45\linewidth]{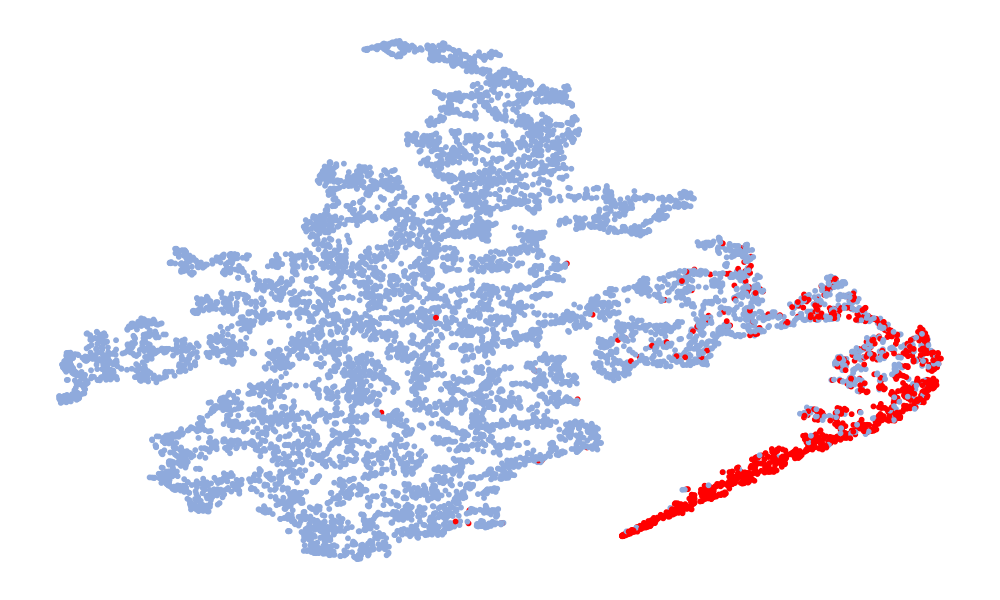}%
		\quad\label{fig_first_case}}
	\subfloat[DKDSSD]{\includegraphics[width=0.45\linewidth]{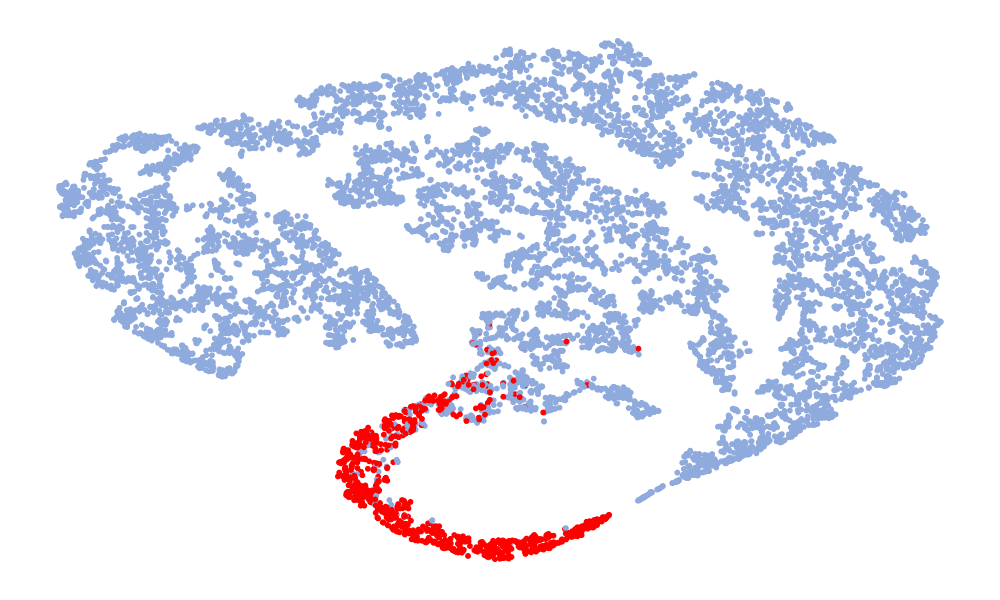}%
		\quad\label{fig_first_case}}
	\\
	\subfloat[Cascade]{\includegraphics[width=0.45\linewidth]{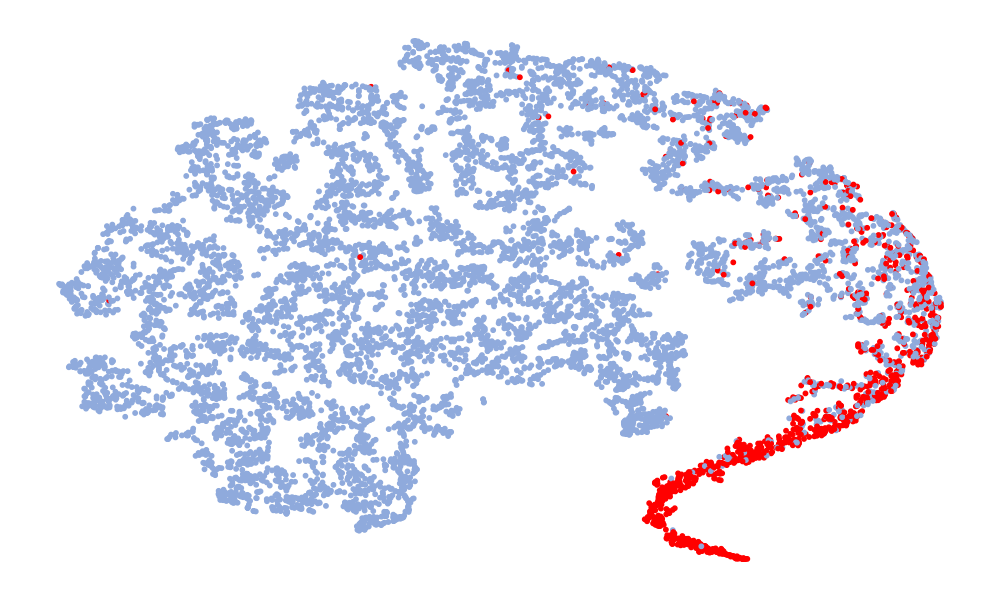}%
		\quad\label{fig_first_case}}
	\subfloat[Cascade]{\includegraphics[width=0.45\linewidth]{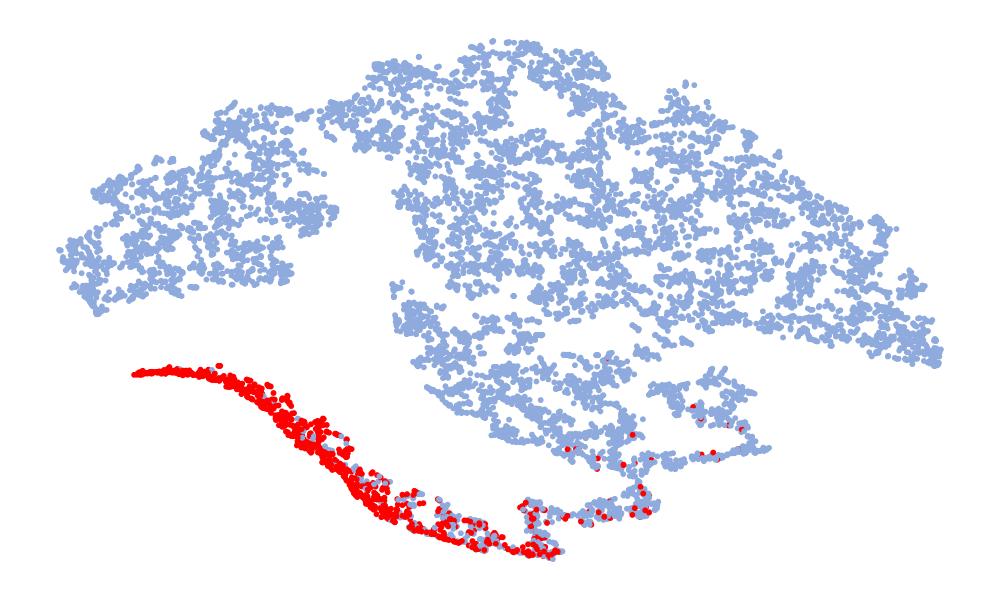}%
		\quad\label{fig_first_case}}
	\\
	\subfloat[Joint]{\includegraphics[width=0.45\linewidth]{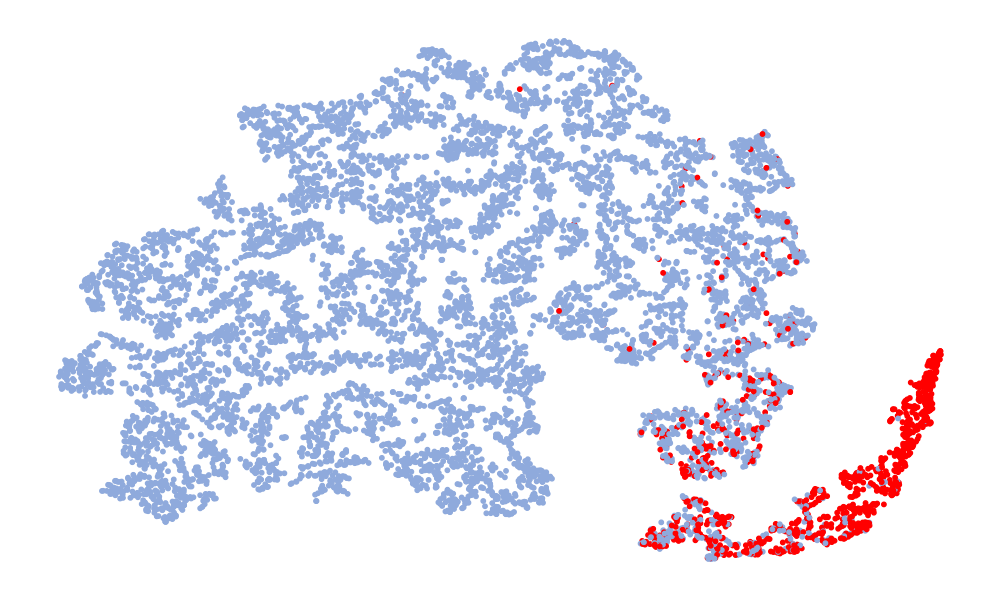}%
		\quad\label{fig_first_case}}
	\subfloat[Joint]{\includegraphics[width=0.45\linewidth]{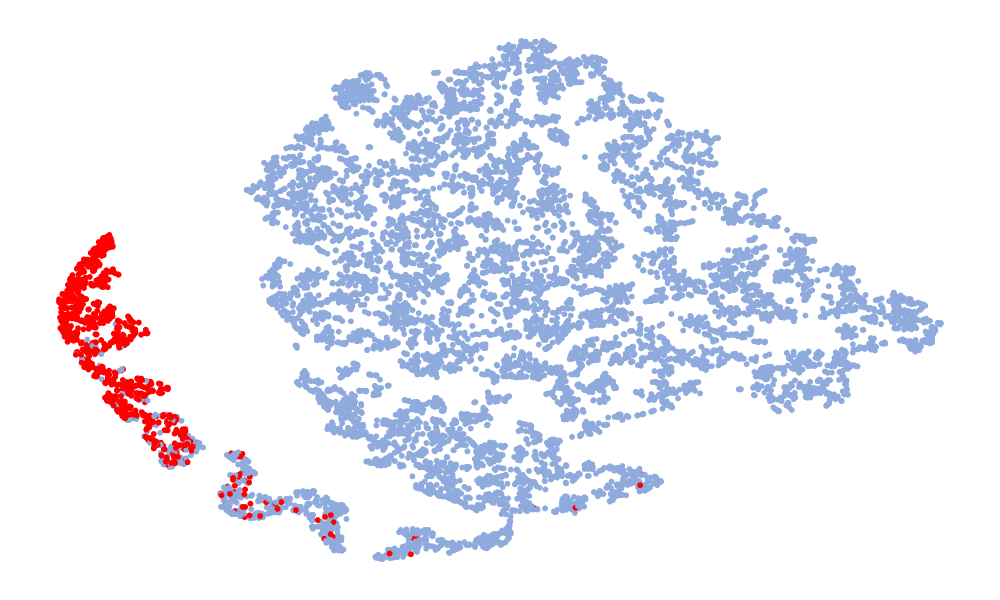}%
		\quad\label{fig_first_case}}
	\\
	\subfloat[MCT1]{\includegraphics[width=0.45\linewidth]{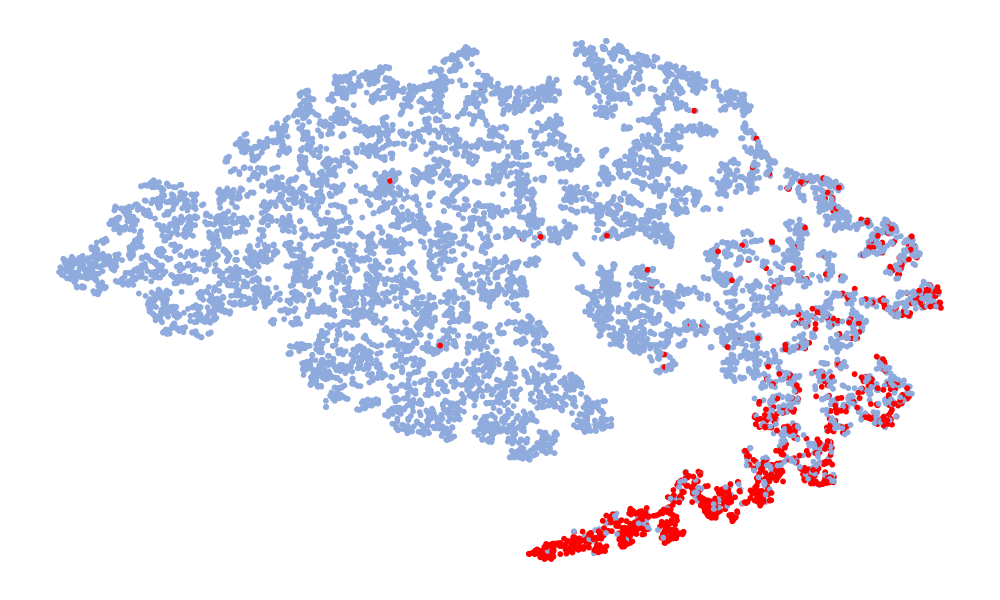}%
		\quad\label{fig_first_case}}
	\subfloat[MCT1]{\includegraphics[width=0.45\linewidth]{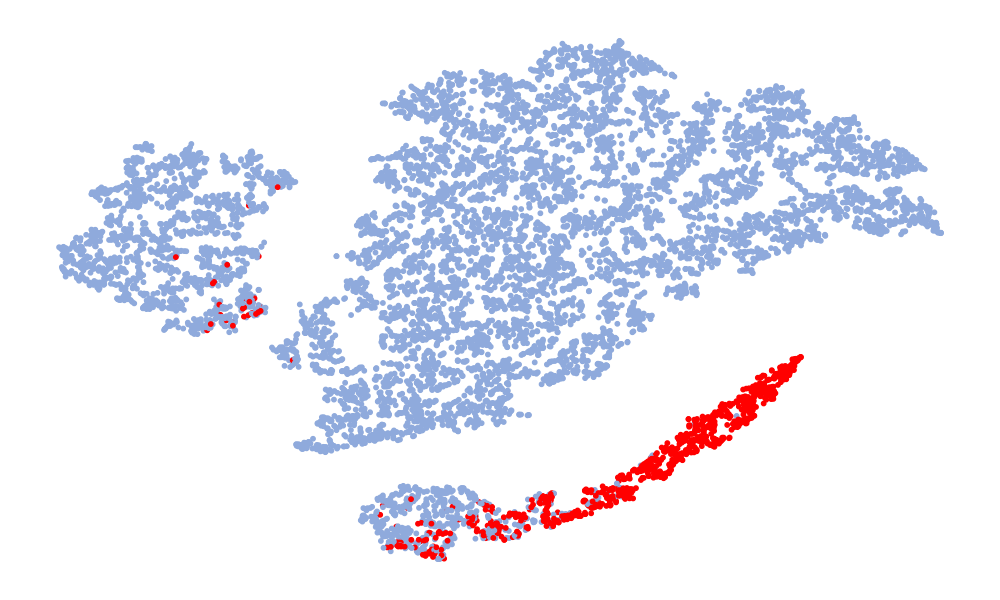}%
		\quad\label{fig_first_case}}
	\\
	\subfloat[MCT2]{\includegraphics[width=0.45\linewidth]{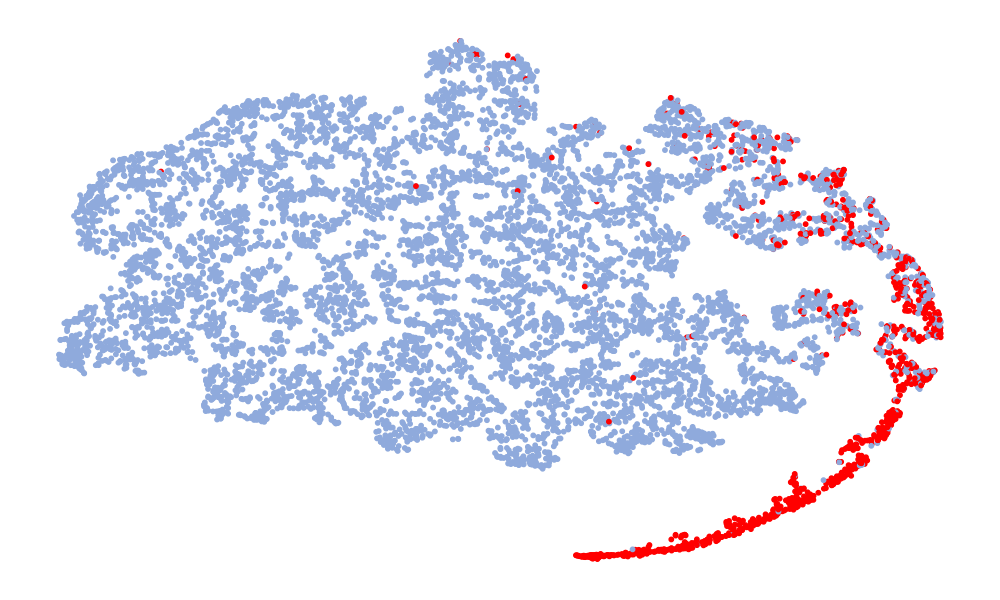}%
		\quad\label{fig_first_case}}
	\subfloat[MCT2]{\includegraphics[width=0.45\linewidth]{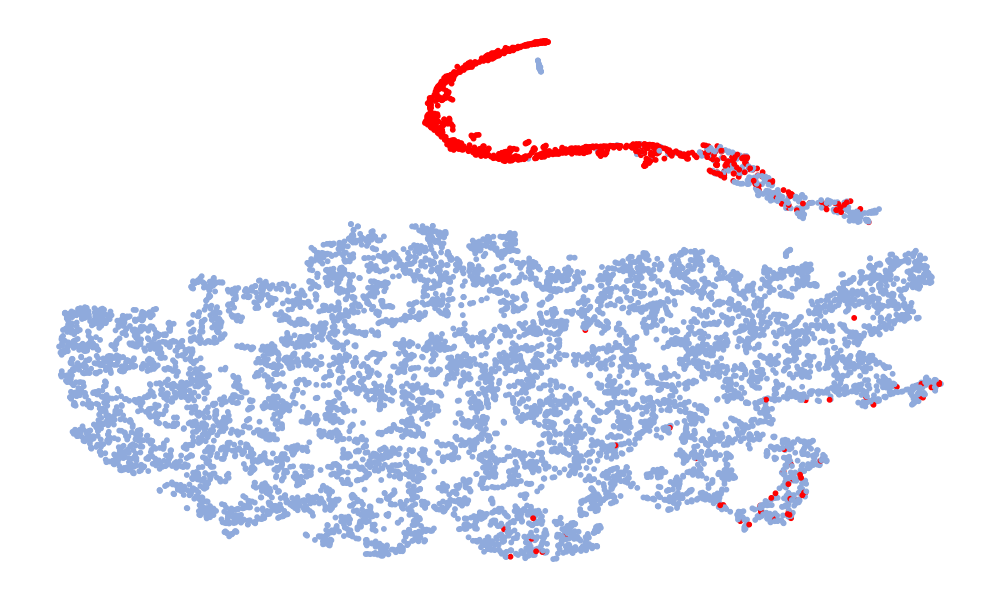}%
		\quad\label{fig_first_case}}
	\\
	\subfloat[Noise-Free]{\includegraphics[width=0.45\linewidth]{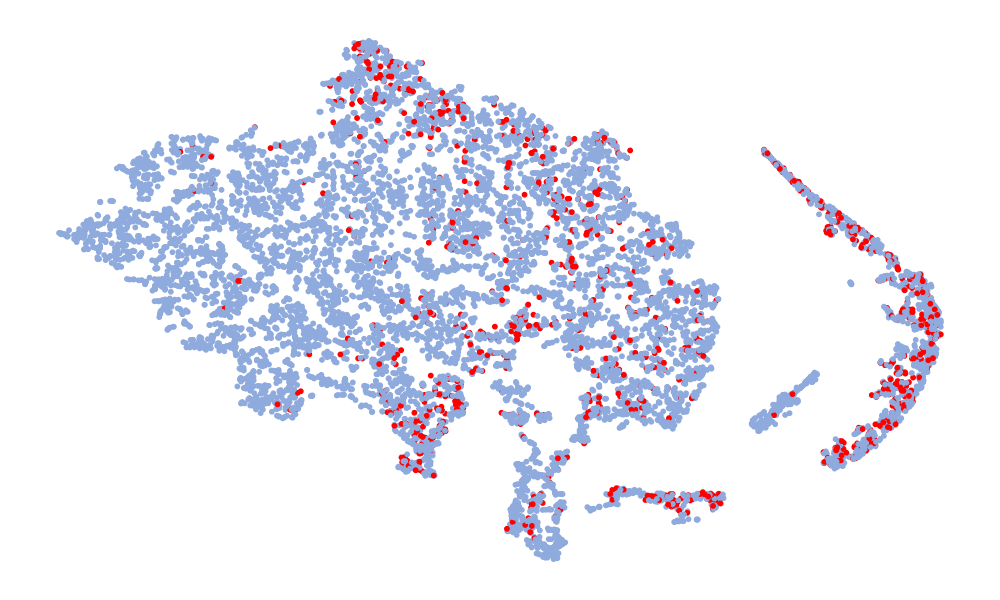}%
		\quad\label{fig_first_case}}
	\subfloat[Noise-Free]{\includegraphics[width=0.45\linewidth]{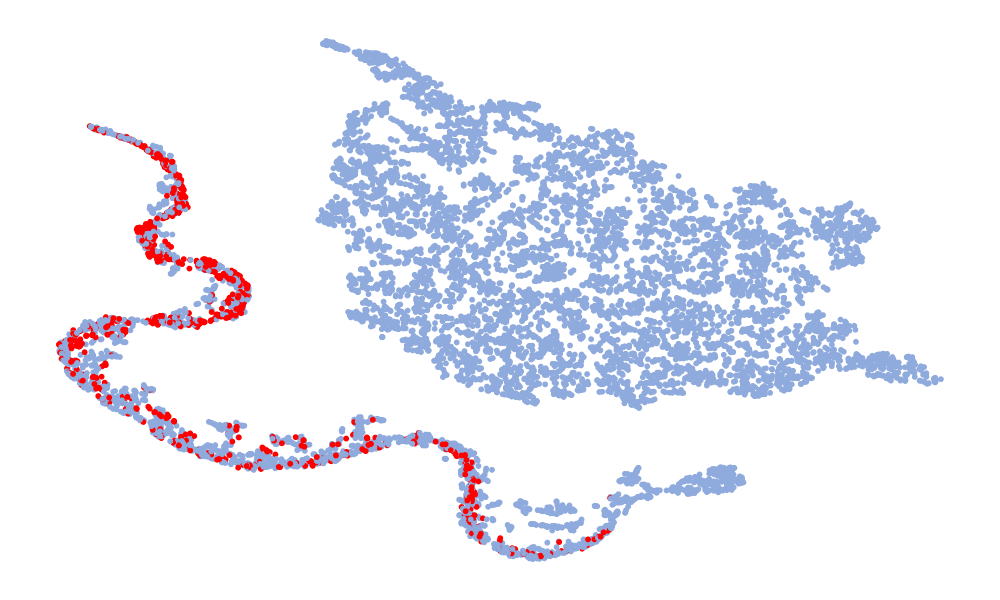}%
		\quad\label{fig_first_case}}
	
	\caption{The t-SNE plots of all baseline models on the ASVspoof 2019 unseen test set (left column) and ASVspoof 2021 LA test set (right column).}
	\label{fig_SNE_PCA}
\end{figure}

\textbf{Impact of Pre-Trained SE Model.} Diffusion is a diffusion-based generative model trained on the WSJ0 \cite{garofalo2007csr}, CHiME3 \cite{barker2015third}, and VoiceBank-DEMAND \cite{valentini2016investigating}. It performs best on datasets containing real noise. We used the pre-trained model of Diffusion for speech enhancement to test whether mature pre-trained models are more advantageous for downstream tasks. From the Table~\ref{tab:newablation}, it can be observed that ``Diffusion+SeNet" performs consistently poorly under all SNR conditions, with results hovering around 20\%. Upon listening to the denoised speech by the diffusion model, we found its speech enhancement effect to be superior, effectively filtering out almost all noise. However, due to the lack of interaction between the pre-trained model and the downstream detection task, there exists an issue of excessive denoising. Speech distortion complicates the distinction between real and fake features, likely contributing to the poor performance. Incorporating the interactive fusion module (``Diffusion+IF+SeNet") lead to a significant performance improvement, reducing the average EER from 21.75\% to 8.62\%. We chose to train a speech enhancement model instead of using an existing pre-trained model. Because for downstream tasks, what is needed is to preserve the effective features of artifacts, not just reduce noise. Although pre-trained speech enhancement front-ends demonstrate strong performance, they require joint optimization with downstream tasks for effective spoofing detection. Other studies, such as \cite{wang2023robust}, have also shown that a pre-trained frontend may not always be the optimal solution. From a parameter perspective, pre-trained models typically result in larger model sizes, which can significantly slow down inference speed and pose challenges for deployment.

\subsection{Visual Analysis}
First, we used t-distributed Stochastic Neighbor Embedding (t-SNE) to reduce the dimensionality of the last layer features of all 6 models and visualize the feature distribution on both the ASVspoof2019 unseen dataset and ASVspoof2021 dataset. Secondly, to analyze the role of the interactive fusion module, we visualized the spectrum features before and after fusion, as well as the fusion weight matrix learned by the interactive fusion module. Finally, statistical analysis was performed on the fusion weight matrix.

\subsubsection{Visualization of Feature Embeddings}
Fig.~\ref{fig_SNE_PCA} is a visualization of features, and the features taken are 128-dimensional features before classification. The red in the figure represents the real speech, and the blue represents the fake speech. The left column is the t-SNE of the 6 models on the ASVspoof 2019 unseen data set, and the right column is the t-SNE of the model on the ASVspoof 2021. Considering the Noise-Free model (Fig.~\ref{fig_SNE_PCA} (f), (l)), it can be seen that the feature spaces of the real speech and the fake speech overlap, and the noise seriously interferes with the distribution of the data. On the whole, in the t-SNE visualization, the distribution of real data appears relatively concentrated, whereas the distribution of fake data appears more scattered due to various types of attacks. Although the feature embedding distributions of different models for real speech are similar, the distribution for fake speech varies considerably. This difference may be attributed to noise interference, which can obscure artifacts and alter the feature distribution. Compared to other models, the t-SNE visualization of the ``DKDSSD" model shows clearer separation between red and blue clusters, indicating stronger capabilities in real and fake classification.

\begin{figure}[t]
	\centering
	\setlength{\abovecaptionskip}{0ex}
	\setlength{\belowcaptionskip}{-10ex}
	\includegraphics[width=\linewidth]{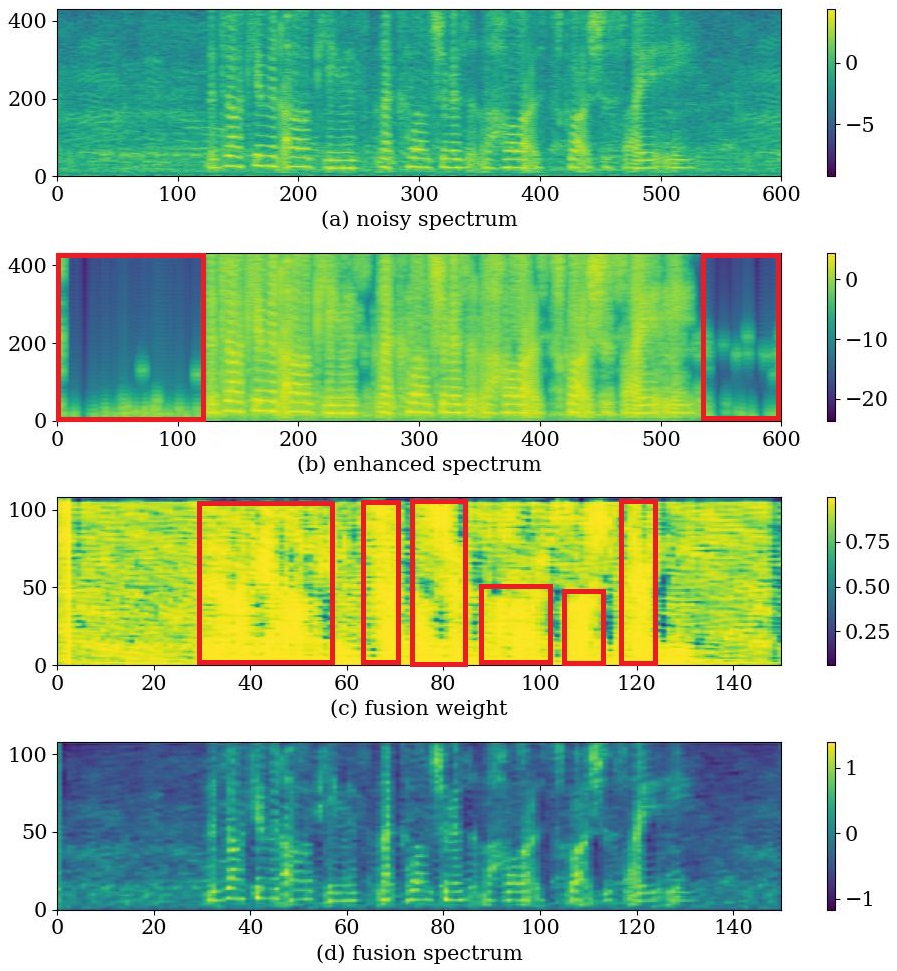}
	\caption{A feature visualization of LA\_E\_2178426\_snr15\_babble.wav. This speech is generated by LA\_E\_2178426.wav in ASVspoof 2019 by superimposing babble noise with an SNR of 15dB. From top to bottom are the logarithmic magnitude spectrum of the noisy speech and the enhanced speech, the fusion weight matrix, and the mean value of the fused 16-channel features.
}
	\label{fig:speech}
\end{figure}

\subsubsection{Visualization of Features and Weight of Interactive Fusion}
Fig.~\ref{fig:speech} is a feature visualization of a single speech (LA\_E\_2178426\_snr15\_babble.wav). This speech is generated by LA\_E\_2178426.wav in ASVspoof 2019 by superimposing destroyer noise with an SNR of 15dB. The number of channels of the feature after the interactive fusion is 16, and the fourth line of the figure is obtained after averaging the dimension of the channel. It can be observed that in areas with human voices (the red box part in Fig.~\ref{fig:speech} (c)), the noisy spectrum occupies a larger weight, while in areas without human voices, the enhanced spectrum has a larger weight. This suggests that the interactive fusion module tends to absorb vocal segments in noisy speech and retain more enhanced features in other areas. In the enhanced features, there is almost no noise in the silent segment (the red box in Fig.~\ref{fig:speech} (b)). After interactive fusion, the speech features (Fig.~\ref{fig:speech} (d)) present a clearer structure than the noisy spectrum, are less affected by noise, and reduce speech distortion. These results demonstrate that the interactive fusion module can adaptively absorb human speech segments while preserving the less noise-contaminated state of enhanced features.

\subsubsection{Visualization of Fusion Weight Statistics for Interactive Fusion} 
\begin{figure}[t]
	\centering
	\setlength{\abovecaptionskip}{0ex}
	\setlength{\belowcaptionskip}{-10ex}
	\includegraphics[width=\linewidth]{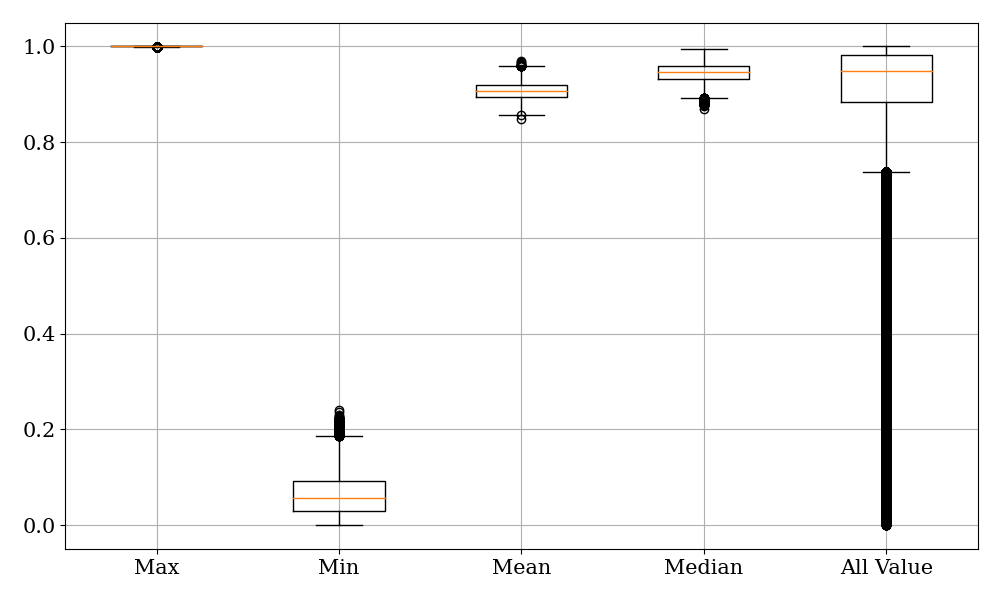}
	\caption{Boxplot of statistical values for interactive fusion weight matrices. From left to right are the maximum value, minimum value, mean value, median value, and all mask values.}
	\label{fig:statics}
\end{figure}

Fig.~\ref{fig:statics} shows box plots of the maximum value, minimum value, mean, median, and all mask values from left to right. It can be observed that the majority of mask values are distributed between 0.8 and 1. These mask values represent the weights occupied by the original noisy features during fusion. It implies that during fusion, the model tends to absorb more of the original noisy spectrum. We speculate that this is because the original speech contains more information and suffers from no distortion issues. The primary role of the enhanced spectrum is to provide information about the non-speech regions. And in a piece of speech, the human voice usually occupies the main part. It confirms that the interactive fusion module tends to absorb human voice segments from the noisy speech and retains more enhanced features in non-speech regions.

\section{Conclusions}
This paper proposes a dual-branch synthetic speech detection method based on knowledge distillation to address the problem of insufficient robustness in noisy scenes. Interactive fusion and knowledge distillation are proposed to guide the training of noisy data, prompting it to behave like clean speech. Interactive fusion adaptively combines noisy data with enhanced data by leveraging channel interaction and spatial information fusion, prompting student branches to generate features that are not disturbed by noise. Knowledge distillation can constrain the decision of the student model through the teacher-student paradigm, mapping the decision space of the noisy student branch to the decision space of the clean teacher branch, allowing students to make final predictions like teachers. Experimental results show that our proposed DKDSSD outperforms other baseline models, especially in low SNR scenarios, and also shows strong generalization in cross-dataset experiments. After visualizing the fusion weight matrices, we observe that noisy speech plays a major role, while the enhanced features contribute information that would otherwise be silent segments. In the future, we will pay more attention to the role of silent segments in detecting synthetic speech.

\bibliographystyle{IEEEtran}
\bibliography{refs}
\begin{IEEEbiography}[{\includegraphics[width=1.1in,height=1.25in,clip,keepaspectratio]{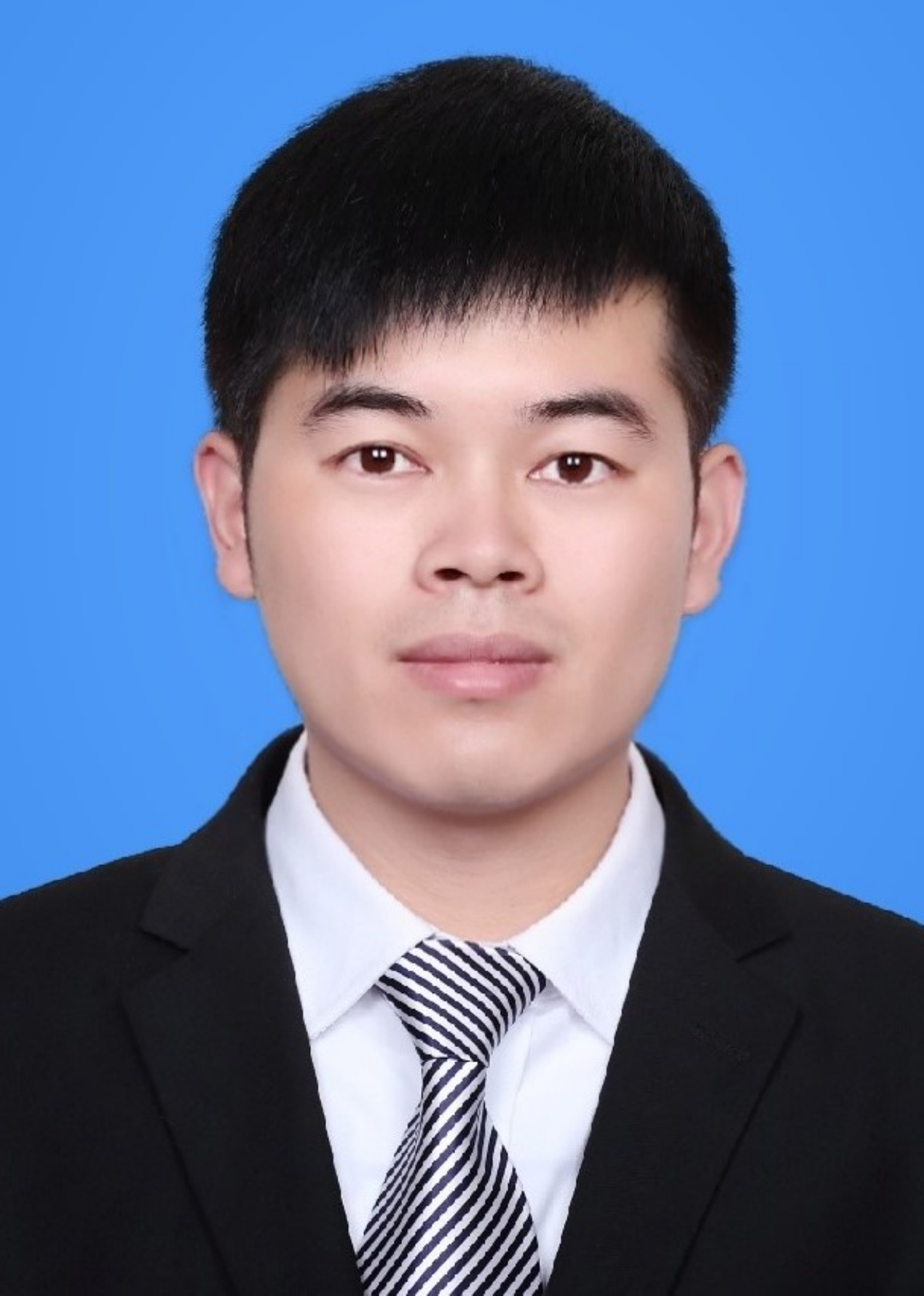}}]{Cunhang Fan} (Member, IEEE)
	received the Ph.D degree with the National Laboratory of Pattern Recognition (NLPR), Institute of Automation, Chinese Academy of Sciences (CASIA), Beijing, China, in 2021, and the B.S. degree from the Beijing University of Chemical Technology (BUCT), Beijing, China, in 2016. He is currently an associate professor with the School of Computer Science and Technology, Anhui University, Heifei, China. His current research interests include speech enhancement, fake speech detection, speech recognition and speech processing.
\end{IEEEbiography}
\vspace{-10 mm} 
\begin{IEEEbiography}[{\includegraphics[width=1in,height=1.25in,clip,keepaspectratio]{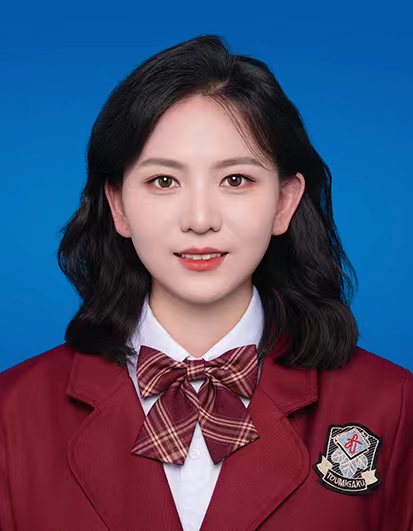}}]{Mingming Ding}  received a B.S. degree in civil engineering from Anhui Jianzhu University in 2018. She is currently currently studying for a M.S. degree in computer science and technology from Anhui University. Her main interests include speech enhancement, speech anti-spoofing and deep learning.
\end{IEEEbiography}
\vspace{-10 mm}

\begin{IEEEbiography}[{\includegraphics[width=1.1in,height=1.25in,clip,keepaspectratio]{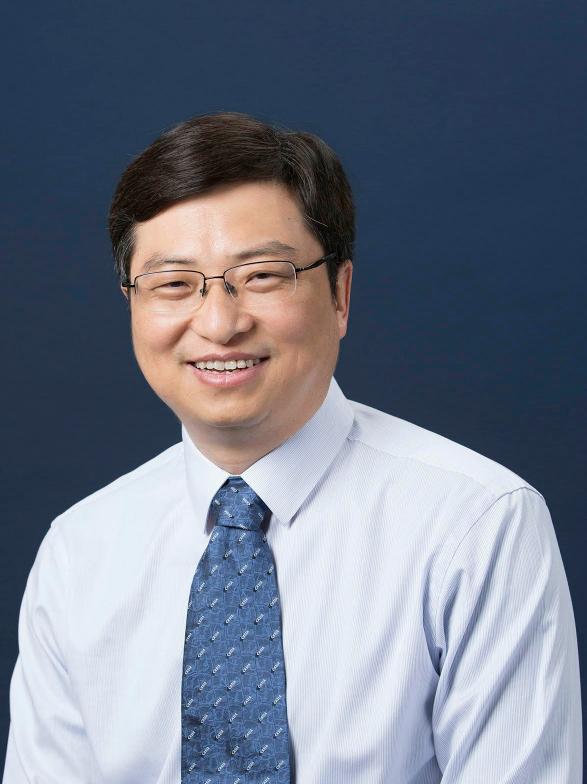}}]{Jianhua Tao}
	(Senior Member, IEEE) received the M.S. degree from Nanjing University, Nanjing, China, in 1996, and the Ph.D. degree from Tsinghua University, Beijing, China, in 2001. He is currently a Professor with Department of Automation, Tsinghua University, Beijing, China. He has authored or coauthored more than 300 papers on major journals and proceedings including the IEEE TASLP, IEEE TAFFC, IEEE TIP, IEEE TSMCB, Information Fusion, etc. His current research interests include speech recognition and synthesis, affective computing, and pattern recognition. He is the Board Member of ISCA, the chairperson of ISCA SIG-CSLP, the Chair or Program Committee Member for several major conferences, including Interspeech, ICPR, ACII, ICMI, ISCSLP, etc. He was the subject editor for the Speech Communication, and is an Associate Editor for Journal on Multimodal User Interface and International Journal on Synthetic Emotions. He was the recipient of several awards from important conferences, including Interspeech, NCMMSC, etc.
\end{IEEEbiography}
\vspace{-10 mm}

\begin{IEEEbiography}[{\includegraphics[width=1in,height=1.25in,clip,keepaspectratio]{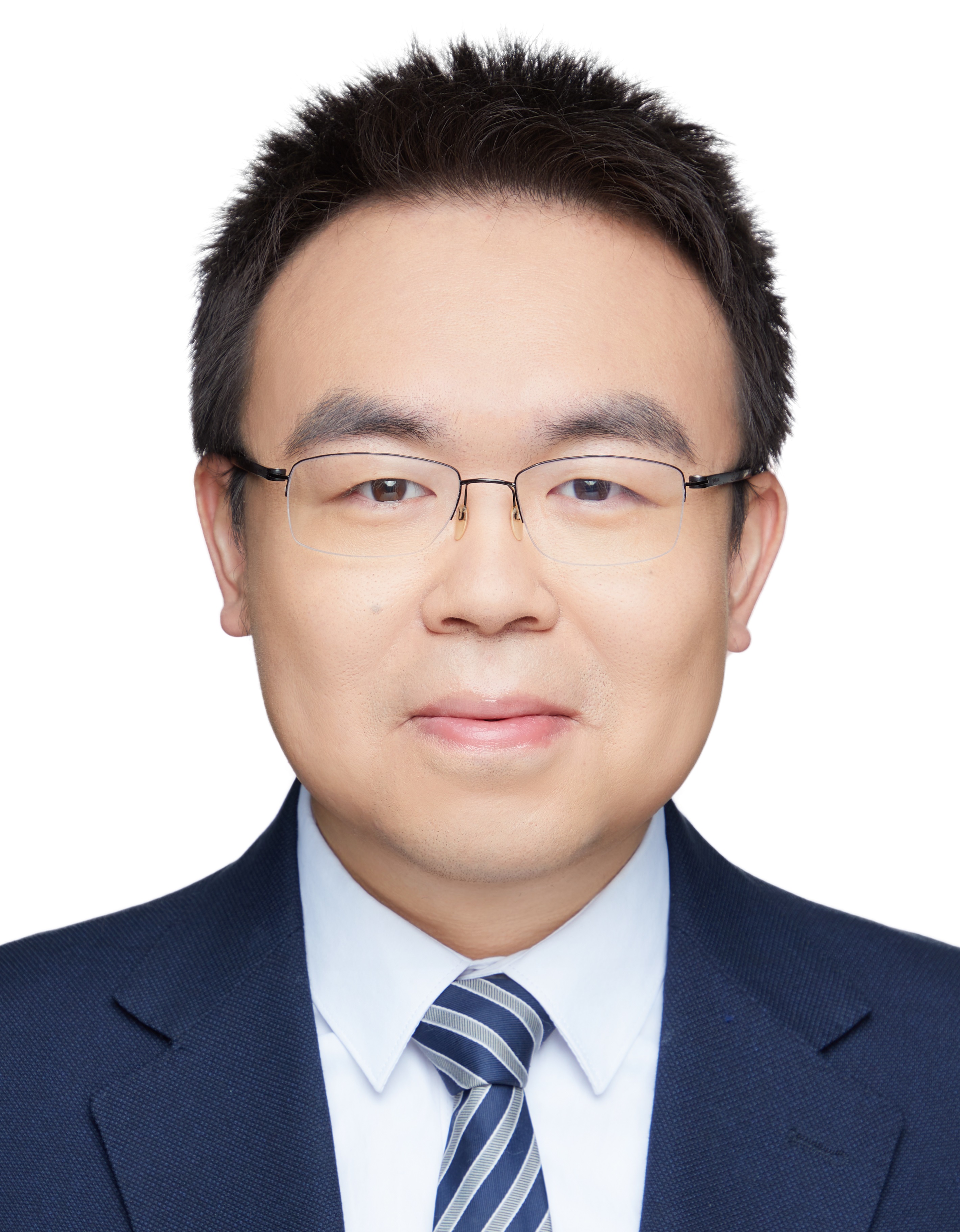}}]{Ruibo Fu} (Member, IEEE) is an assistant professor in the National
	Laboratory of Pattern Recognition, Institute of Automation, Chinese Academy
	of Sciences, Beijing. He obtained B.E. from Beijing University of Aeronautics and Astronautics in 2015 and Ph.D. from the Institute of Automation, Chinese Academy of Sciences in 2020. His research interest is speech synthesis and transfer learning. He has published more than 20 papers in international conferences and journals such as ICASSP and INTERSPEECH and has won the best paper award twice in NCMMSC 2017 and 2019. He won the first prize in the personalized speech synthesis competition held by the Ministry of Industry and Information Technology twice in 2019 and 2020. He also won the first prize in the ICASSP2021 Multi-Speaker Multi-Style Voice Cloning Challenge (M2VoC) Challenge.
\end{IEEEbiography}
\vspace{-10 mm} 
\begin{IEEEbiography}[{\includegraphics[width=1.1in,height=1.25in,clip,keepaspectratio]{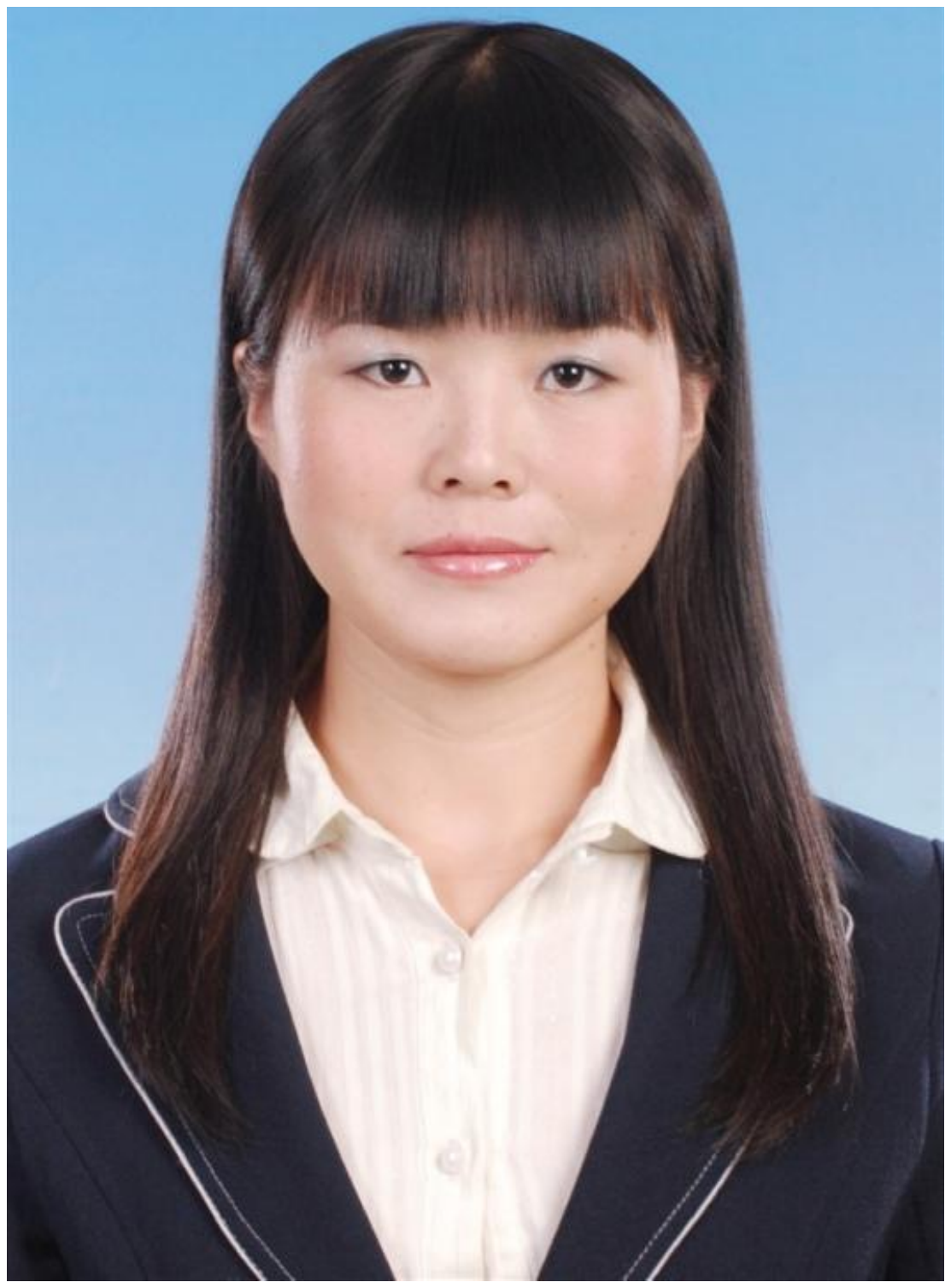}}]{Jiangyan Yi} (Member, IEEE) received the Ph.D. degree from the University of Chinese Academy of Sciences, Beijing, China, in 2018, and the M.A. degree from the Graduate School of Chinese Academy of Social Sciences, Beijing, in 2010. During 2011 to 2014, she was a Senior R\&D Engineer with Alibaba Group. She is currently an Associate Professor with the State Key Laboratory of Multimodal Artificial Intelligence Systems, Institute of Automation, Chinese Academy of Sciences. Her research interests include speech signal processing, speech recognition and synthesis, fake audio detection, audio forensics, and transfer learning.
\end{IEEEbiography}
\vspace{-10 mm} 

\begin{IEEEbiography}[{\includegraphics[width=1.1in,height=1.25in,clip,keepaspectratio]{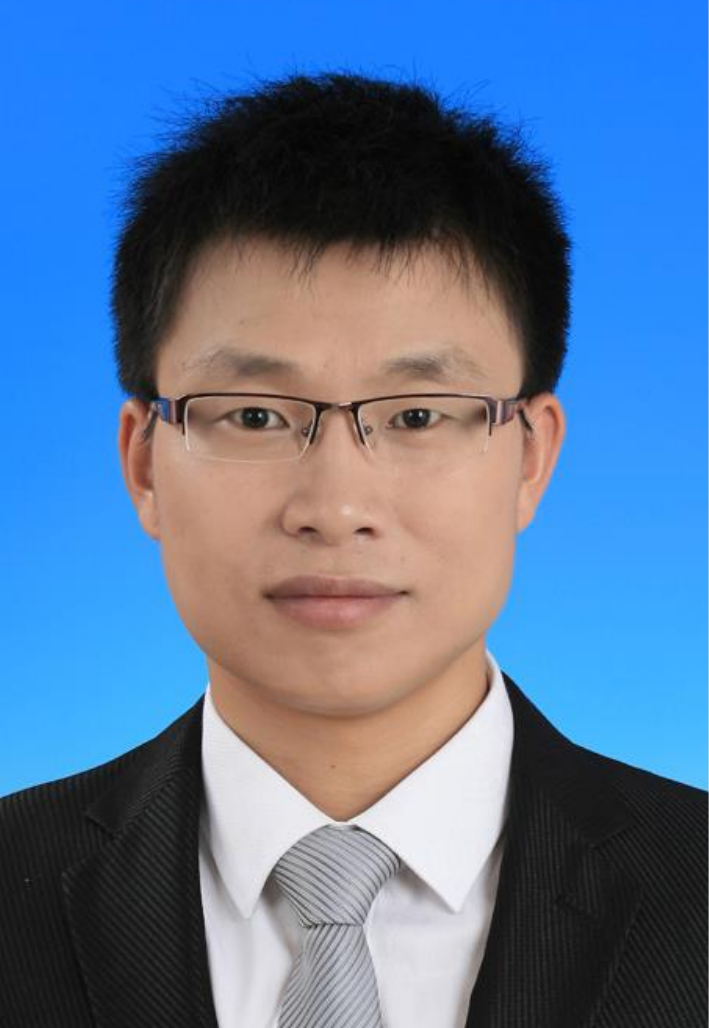}}]{Zhengqi Wen} (Member, IEEE) received the B.S.degree from the University of Science and Technology of China, Hefei, China, in 2008, and the Ph.D. degree from the Chinese Academy of Sciences, Beijing, China, in 2013, both in pattern recognition and intelligent system. He is currently an Associate Professor with the National Laboratory of Pattern Recognition Institute of Automation, Chinese Academy of Sciences. His current research interests include speech processing, speech recognition, and speech synthesis.
\end{IEEEbiography}
\vspace{-10 mm} 

\begin{IEEEbiography}[{\includegraphics[width=1.1in,height=1.25in,clip,keepaspectratio]{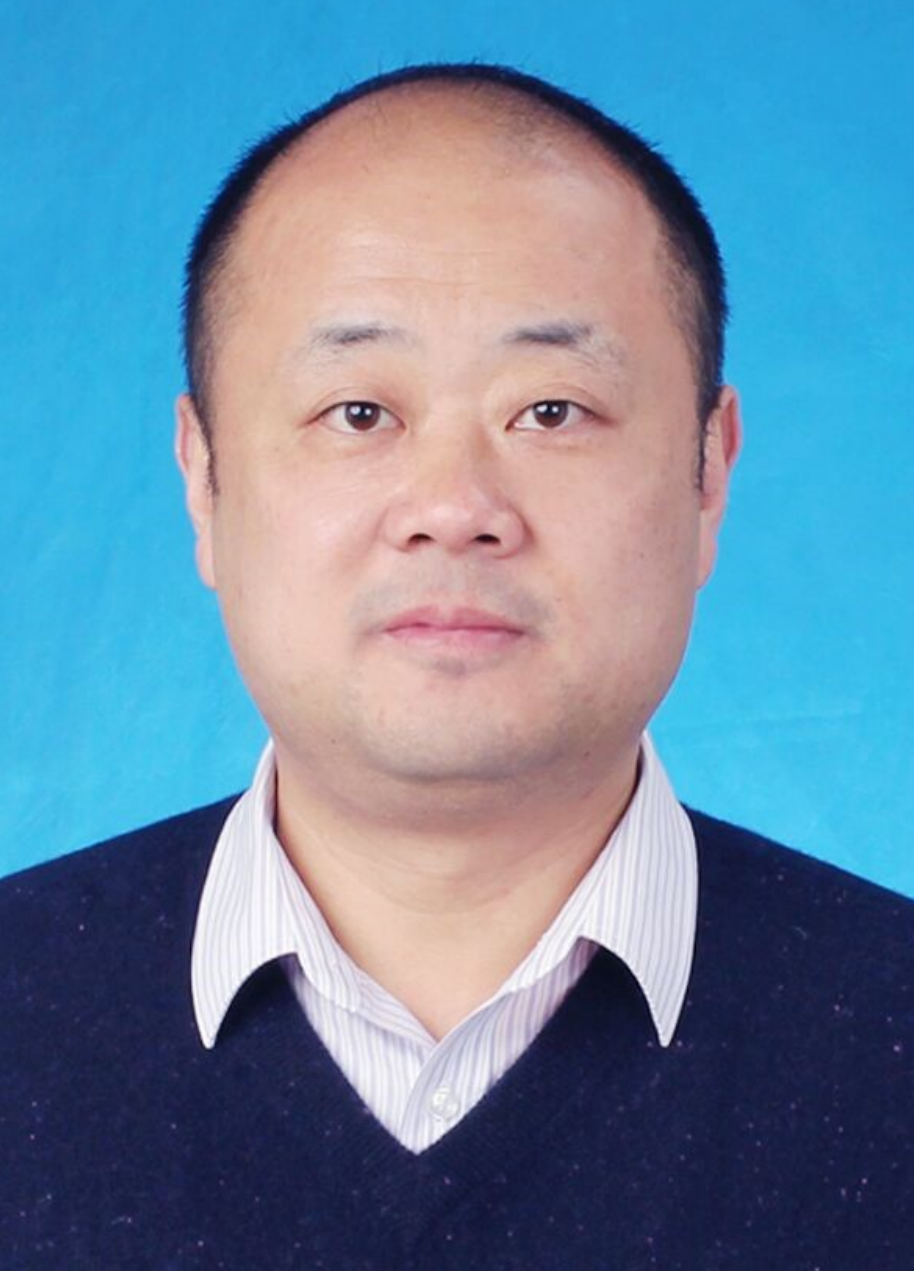}}]{Zhao Lv} (Member, IEEE)
	received his Ph.D. degree in Computer Application Technology from Anhui University, Hefei, China, in 2011. He was a visiting scholar with the University of Utah, Salt Lake City, USA, from 2017 to 2018. He is currently a professor in the School of Computer Science and Technology at Anhui University, Hefei, China. His research interests include intelligent information processing and pattern recognition regarding biomedical signals (EEG, EOG, etc.) as well as speech signal processing.
\end{IEEEbiography}
%


\vfill

\end{document}